\documentclass{pasj00}
\usepackage{natbib}
\bibpunct{(}{)}{;}{a}{}{,}

\begin{document}
\SetRunningHead{}{}

\title{SGRs and AXPs as rotation powered massive white dwarfs}

\author{
Manuel \textsc{Malheiro}\altaffilmark{1,4}
Jorge A. \textsc{Rueda}\altaffilmark{1,2}
and
Remo \textsc{Ruffini}\altaffilmark{1,2,3}
\thanks{email:ruffini@icra.it}
}
\altaffiltext{1}{Dipartimento di Fisica and ICRA, Sapienza Universit\`a di Roma, P.le Aldo Moro 5, I--00185 Rome, Italy}
\altaffiltext{2}{ICRANet, P.zza della Repubblica 10, I--65122 Pescara, Italy}
\altaffiltext{3}{ICRANet, University of Nice-Sophia Antipolis, 28 Av. de Valrose, 06103 Nice Cedex 2, France}
\altaffiltext{4}{Departamento de F\'{i}sica, Instituto Tecnol\'{o}gico de Aeron\'{a}utica, CTA, S\~{a}o Jos\'{e} dos Campos, 12.228-900, Brazil}



\maketitle

\begin{abstract}
SGR 0418+5729 is a ``Rosetta Stone'' for deciphering the energy source of Soft Gamma Ray Repeaters (SGRs) and Anomalous X-ray Pulsars (AXPs). We show a model based on canonical physics and astrophysics for SGRs and AXPs powered by massive highly magnetized rotating white dwarfs (WDs), in total analogy with pulsars powered by rotating neutron stars (NSs). We predict for SGR 0418+5729 a lower limit for its spin-down rate, $\dot{P} \geq L_X P^3/(4\pi^2 I)=1.18\times 10^{-16}$ where $I$ is the moment of inertia of the WD. We show for SGRs and AXPs that, the occurrence of the glitch and the gain of rotational energy, is due to the release of gravitational energy associated to the contraction and decrease of the moment of inertia of the WDs. The steady emission and the outburst following the glitch are explained by the loss of rotational energy of the Wds, in view of the much larger moment of inertia of the WDs, as compared to the one of NSs and/or quark stars. There is no need here to invoke the unorthodox concept of magnetic energy release due to decay of overcritical magnetic fields, as assumed in the magnetar model. A new astrophysical scenario for the SGRs and AXPs associated to Supernova remnants is presented. The observational campaigns of the X-ray Japanese satellite Suzaku on AE Aquarii and the corresponding theoretical works by Japanese groups and recent results of the Hubble Space Telescope, give crucial information for our theoretical model. Follow-on missions of Hubble Telescope and VLT are highly recommended to give further observational evidence of this most fundamental issue of relativistic astrophysics: the identification of the true SGRs/AXPs energy source.
\end{abstract}

\section{Introduction}\label{sec:1}

Soft Gamma Ray Repeaters (SGRs) and Anomalous X-ray Pulsars (AXPs) are a class of compact objects that show interesting observational properties \citep[see e.g.][]{mereghetti08}: rotational periods in the range $P\sim (2$--$12)$ s, a narrow range with respect to the wide range of ordinary pulsars $P\sim (0.001$--$10)$ s; spin-down rates $\dot{P} \sim (10^{-13}$--$10^{-10})$, larger than ordinary pulsars $\dot{P} \sim 10^{-15}$; strong outburst of energies $\sim (10^{41}$--$10^{43})$ erg, and for the case of SGRs, giant flares of even large energies $\sim (10^{44}$--$10^{47})$ erg, not observed in ordinary pulsars.

The recent observation of SGR 0418+5729 with a rotational period of $P=9.08$ s, an upper limit of the first time derivative of the rotational period $\dot{P} < 6.0 \times 10^{-15}$ \citep{rea10}, and an X-ray luminosity of $L_X = 6.2\times 10^{31}$ erg/s promises to be an authentic Rosetta Stone, a powerful discriminant for alternative models of SGRs and AXPs. 

If described as a neutron star of $M=1.4 M_\odot$, $R=10$ km and a moment of inertia $I\approx 10^{45}$ g cm$^2$, which we adopt hereafter as fiducial parameters, the loss of rotational energy of the neutron star 
\begin{equation}\label{eq:EdotNS}
\dot{E}^{\rm NS}_{\rm rot}=-4 \pi^2 I \frac{\dot{P}}{P^3} = -3.95\times 10^{46} \frac{\dot{P}}{P^3}\quad {\rm erg/s}\, ,
\end{equation}
associated to its spin-down rate $\dot{P}$, cannot explain the X-ray luminosity of SGR 0418+5729, i.e. $\dot{E}^{\rm NS}_{\rm rot}<L_X$, excluding the possibility of identifying this source as an ordinary spin-down powered pulsar. 

The magnetar model of SGRs and AXPs, based on a neutron star of fiducial parameters, needs a magnetic field larger than the critical field for vacuum polarization $B_c=m^2_e c^3/(e \hbar)=4.4\times 10^{13}$ G in order to explain the observed X-ray luminosity in terms of the release of magnetic energy \citep[see][for details]{duncan92,thompson95}. However, the inferred upper limit of the surface magnetic field of SGR 0418+5729 $B<7.5\times 10^{12}$ G describing it as a neutron star \citep[see][for details]{rea10}, is well below the critical field challenging the power mechanism based on magnetic field decay purported in the magnetar scenario.

We show that the observed upper limit on the spin-down rate of SGR 0418+5729 is, instead, perfectly in line with a model based on a massive fast rotating highly magnetized white dwarf \citep[see e.g.][]{paczynski90} of mass $M=1.4M_\odot$, radius $R=10^3$ km, and moment of inertia $I\approx 10^{49}$ g cm$^2$, which we adopt hereafter as fiducial white dwarf parameters. Such a configuration leads for SGR 0418+5729 to a magnetic field $B < 7.5 \times 10^8$ G. The X-ray luminosity can then be expressed as originating from the loss of rotational energy of the white dwarf leading to a theoretical prediction for the first time derivative of the rotational period
\begin{equation}\label{eq:Pdotnew}
\frac{L_X P^3}{4\pi^2 I} \leq \dot{P}_{\rm SGR 0418+5729} < 6.0\times 10^{-15} \, ,
\end{equation}
where the lower limit is established by assuming that the observed X-ray luminosity of SGR 0418+5729 coincides with the rotational energy loss of the white dwarf. For this specific source, the lower limit of $\dot{P}$ given by Eq.~(\ref{eq:Pdotnew}) is $\dot{P}_{\rm SGR 0418+5729} \geq 1.18\times 10^{-16}$. This prediction is left to be verified by the dedicated scientific missions.

The assumption of massive fast rotating highly magnetized white dwarfs appears to be very appropriate since their observation has been solidly confirmed in the last years thanks to observational campaigns carried out by the X-ray Japanese satellite Suzaku \citep[see e.g.][]{terada08,2008AstHe.101..526T,2008AdSpR..41..512T,2008AIPC.1085..689T,2008HEAD...10.1003T}. The magnetic fields observed in white dwarfs are larger than $10^6$ G all the way up to $10^9$ G \citep[see e.g][]{angel81,ferrario97,nalezyty04,ferrario05,terada08,2009A&A...506.1341K}. These observed massive fast rotating highly magnetized white dwarfs share common properties with SGRs/AXPs. The specific comparison between SGR 0418+5729 and the white dwarf AE Aquarii \citep{terada08} is given in Sec.~\ref{sec:4}.

The aim of this article is to investigate the implications of the above considerations to all observed SGRs and AXPs. The article is organized as follows. In Sec.~\ref{sec:2} we summarize the main features of a model for SGRs and AXPs based on rotation powered white dwarfs while, in Sec.~\ref{sec:3}, we recall the magnetar model. In Sec.~\ref{sec:4} we present the observations of massive fast rotating highly magnetized white dwarfs. The constraints on the rotation rate imposed by the rotational instabilities of fast rotating white dwarfs are discussed in Sec.~\ref{sec:5} and in Sec.~\ref{sec:6} we analyze the glitch-outburst connection in SGRs and AXPs. The magnetospheric emission from the white dwarf is discussed in Sec.~\ref{sec:7} and the possible connection between SGRs and AXPs with supernova remnants is presented in Sec.~\ref{sec:8}. In Sec.~\ref{sec:9} we address the problem of fiducial parameters of both white dwarfs and neutron stars and, in Sec.~\ref{sec:10}, we summarize conclusions and remarks.

\section{SGRs and AXPs within the white dwarf model}\label{sec:2}

We first recall the pioneering works of \cite{1988ApJ...333..777M} and \cite{paczynski90} on 1E 2259+586. This source is pulsating in the X-rays with a period $P=6.98$ s \citep{fahlman81}, a spin-down rate of $\dot{P}=4.8 \times 10^{-13}$ \citep{davies90} and X-ray luminosity $L_X = 1.8 \times 10^{34}$ erg/s \citep{gregory80,hughes81,1988ApJ...333..777M}. Specially relevant in the case of 1E 2259+586 is also its position within the supernova remnant G109.1-1.0 with age estimated $t-t_0=(12$--$17)$ kyr \citep{gregory80,hughes81}.

Paczynski developed for 1E 2259+586 a model based on a massive fast rotating highly magnetized white dwarf. The upper limit on the magnetic field \citep[see e.g.][]{ferrari69} obtained by requesting that the rotational energy loss due to the dipole field be smaller than the electromagnetic emission of the dipole, is given by
\begin{equation}\label{eq:Bmax}
B=\left( \frac{3 c^3}{8 \pi^2} \frac{I}{R^6} P \dot{P} \right)^{1/2}\, ,
\end{equation}
where $P$ and $\dot{P}$ are observed properties and the moment of inertia $I$ and the radius $R$ of the object are model dependent properties. For the aforementioned fiducial parameters of a fast rotating magnetized white dwarf, Eq.~(\ref{eq:Bmax}) becomes
\begin{equation}\label{eq:BmaxWD}
B=3.2\times 10^{15} \left(P \dot{P} \right)^{1/2} {\rm G}\, .
\end{equation}

The loss of rotational energy within this model is given by
\begin{equation}\label{eq:Edot}
\dot{E}^{\rm WD}_{\rm rot} = - 4 \pi^2 I \frac{\dot{P}}{P^3} = -3.95\times 10^{50} \frac{\dot{P}}{P^3}\quad {\rm erg/s}\, ,
\end{equation}
which amply justifies the steady X-ray emission of 1E 2259+586 (see Table \ref{tab:properties}).

A further development for the source 1E 2259+586, came from \cite{usov94}, who introduced the possibility in a white dwarf close to the critical mass limit, to observe sudden changes in the period of rotation, namely glitches. When the rotation of the white dwarf slows down, centrifugal forces of the core decrease and gravity pulls it to a less oblate shape thereby stressing it. The release of such stresses leads to a sudden decrease of moment of inertia and correspondingly, by conservation of angular momentum
\begin{equation}
J= I \Omega = (I+\Delta I)(\Omega+\Delta \Omega)={\rm constant}\, ,
\end{equation}
to a shortening of the rotational period and a shrinking of the stellar radius
\begin{equation}\label{eq:changeMI}
\frac{\Delta I}{I}=2\frac{\Delta R}{R}=\frac{\Delta P}{P}=-\frac{\Delta \Omega}{\Omega}\, ,
\end{equation}
that leads to a change of gravitational energy
\begin{equation}\label{eq:DeltaEg}
\Delta E_{\rm g} = \frac{G M^2}{R}\frac{\Delta R}{R} \sim 2.5\times 10^{51}\frac{\Delta P}{P}\quad{\rm erg}\, ,
\end{equation}
which apply as well in the case of solid quark stars \citep[see e.g.][]{xu06,tong11}.

The fractional change of period (\ref{eq:changeMI}) leads to a gain of rotational energy in the spin-up process of the glitch
\begin{equation}\label{eq:deltaEglitch}
\Delta E^{\rm WD}_{\rm rot}=-\frac{2 \pi^2 I}{P^2}\frac{\Delta P}{P}=-1.98\times 10^{50}\frac{\Delta P}{P^3}\quad {\rm erg}\, ,
\end{equation}
which is fully explained by the available gravitational energy given by Eq.~(\ref{eq:DeltaEg}).

\begin{figure}[h]
\centering
\includegraphics[width=\hsize,clip]{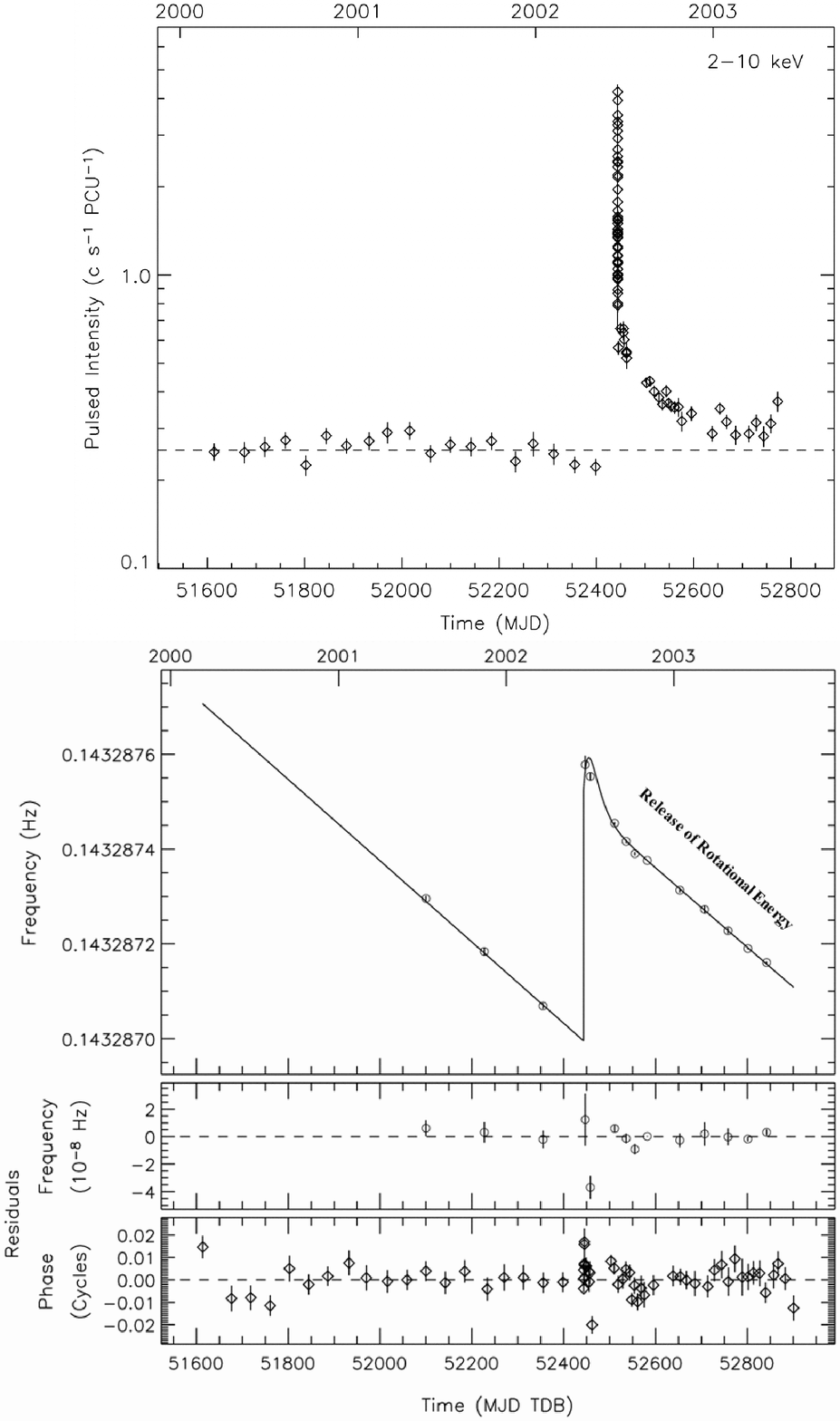}
\caption{Timing and pulsed emission analysis of the glitch-outburst of 1E 2259+586 on June 2002 \citep[taken from][]{woods04}. The observed fractional change of period is $\Delta P/P = - \Delta \Omega/\Omega \sim -4 \times 10^{-6}$ and the observed energy released during the event is $\sim 3\times 10^{41}$ erg \citep{woods04}. Within the white dwarf model from such a $\Delta P/P$ we obtain $\Delta E^{\rm WD}_{\rm rot} \sim 1.7\times 10^{43}$ erg as given by Eq.~(\ref{eq:deltaEglitch}), see Sec.~\ref{sec:6} for details. We have modified the original Fig.~7 of \cite{woods04} by indicating explicitly where the rotational energy is released after the spin-up, during the recovery phase, by the emission of a sequence of bursts on time scales from months to years \citep[see e.g.][]{mereghetti08}.}\label{fig:glitch1E2259}
\end{figure}

If we turn now to the glitch-outburst correlation, the electromagnetic energy of the bursts in the rotating white dwarf model finds its energetic origin in the release of the rotational energy related to the slowing down of the white dwarf rotational frequency. This occurs all the way to the end of the recovery phase, on time scales from months to years. This is most impressively represented e.g. in the case of the glitch-outburst episode of 1E 2259+586 on June 2002 \citep{kaspi03,woods04}, see Fig.~\ref{fig:glitch1E2259} for details. We indeed show in Sec.~\ref{sec:6} that, the change of the moment of inertia of the white dwarf given by Eq.~(\ref{eq:changeMI}), leading to the release of gravitational energy given by Eq.~(\ref{eq:DeltaEg}), and to the rotational energy gain of the white dwarf expressed by Eq.~(\ref{eq:deltaEglitch}), is enough to explain the total electromagnetic energy released in the main burst and in the subsequent activity.

For the evolution of the period close to a glitch we follow the parameterization by \cite{manchesterbook}. The angular velocity $\Omega=2\pi/P$, since the glitch time $t=t_g$, until the complete or partial recovery, can be described as
\begin{equation}
\Omega = \Omega_0(t) + \Delta \Omega [1-Q(1-e^{-(t-t_g)/\tau_d})]\, ,
\end{equation}
where $\Omega_0(t)=\Omega_0 + \dot{\Omega} (t-t_g)$ is the normal evolution of the frequency in absence of glitch, being $\Omega_0$ the frequency prior to the glitch, $\Delta \Omega = -2 \pi \Delta P/P^2$ is the initial frequency jump, which can be decomposed in the persistent and decayed parts, $\Delta \Omega_p$ and $\Delta \Omega_d$ respectively, $\tau_d$ is the timescale of the exponential decay of the frequency after the glitch and $Q=\Delta \Omega_d/\Delta \Omega=1-\Delta \Omega_p/\Delta \Omega$ is the recovery fraction or ``healing parameter''.  For full recovery we have $Q=1$, $\Omega (t>>\tau_d) = \Omega_0$, and for zero recovery $Q=0$, $\Omega (t>>\tau_d) = \Omega_0(t)+\Delta \Omega$. For simplicity we assume in the following and especially below in Sec.~\ref{sec:6}, complete recovery  $Q=1$. 

This mechanism in white dwarfs is similar, although simpler, than the one used to explain e.g. glitches in ordinary pulsars \cite[see e.g.][]{1971AnPhy..66..816B,shapirobook}. The essential difference is that neutron stars are composed by a superfluid core and a solid crust, being the latter the place where starquakes can originate leading to glitches. A two-component description is then needed, see e.g. \cite{shapirobook}. In the present case of a massive rotating white dwarf, such a two-component structure does not exist and the white dwarf behaves as a single solid system. What is important to stress is that the rotational energy released for $Q\geq1$ is largely sufficient for the explanation of the bursting phenomena, see Sec.~\ref{sec:6} for details.

The crystallization temperature of a white dwarf composed of nuclei $(Z,A)$ and mean density $\bar{\rho}$ is given by \citep[see e.g.][]{shapirobook,usov94}
\begin{equation}
T_{\rm cry} \simeq 2.28 \times 10^5 \frac{Z^2}{A^{1/3}} \left( \frac{\bar{\rho}}{10^6 {\rm g/cm}^3} \right)^{1/3} {\rm K}\, .
\end{equation}

Thus, assuming an internal white dwarf temperature $\sim 10^7$ K we find that the mean density for the crystallization of the white dwarf should be $\sim 2.2\times 10^7$ g/cm$^3$ for $^{12}$C, $\sim 5.2\times 10^6$ g/cm$^3$ for $^{16}$O and $\sim 1.25\times 10^6$ g/cm$^3$ for $^{56}$Fe. Very massive white dwarfs as the ones we are considering here have mean densities $\sim 10^9$ g/cm$^3$ and therefore a considerable fraction of their size should be in principle solid at these high temperatures \citep[see also][]{2005A&A...441..689A,2007A&A...465..249A}. It is worth to mention that, the phase separation of the constituents of CO white dwarfs, theoretically expected to occur in the crystallization process \citep[see][for details]{1988Natur.333..642G}, has been recently observationally confirmed solving the puzzle of the age discrepancy of the open cluster NGC 6791 \citep{2010Natur.465..194G}. 

Under these physical conditions, starquakes leading to glitches in the white dwarf may occur with a recurrence time \citep[see e.g.][]{1971AnPhy..66..816B,usov94}
\begin{equation}\label{eq:tq}
\delta t_q = \frac{2 D^2}{B} \frac{|\Delta P|/P}{| \dot{E}_{\rm rot}|}\, ,
\end{equation}
where $\dot{E}_{\rm  rot}$ is the loss of rotational energy (\ref{eq:Edot}), $D =(3/25)\,G M^2_c/R_c$, $B = 0.33\,(4 \pi/3) R^3_c e^2 Z^2 [\bar{\rho}_c/(A m_p)]^{4/3}$, $M_c$, $R_c$ and $\bar{\rho}_c$ are the mass, the radius and the mean density of the solid core, and $m_p$ is the proton mass.

For the specific case of 1E 2259+586, Usov predicted the possible existence of changes of period $\Delta P/P \approx -(1$--$3)\times 10^{-6}$ with a recurrence time between cracks $\delta t_q \approx 7 \times 10^6 \left|\Delta P\right|/P$ yr $\approx$ a few times $(1$--$10)$ yr. It is impressive that in 2002 indeed changes of the order of $\Delta P /P \approx -4 \times 10^{-6}$ were observed in 1E 2259+586 \citep{kaspi03,woods04} (see Fig.~\ref{fig:glitch1E2259} for details).

\begin{figure}[h]
\includegraphics[width=\hsize,clip]{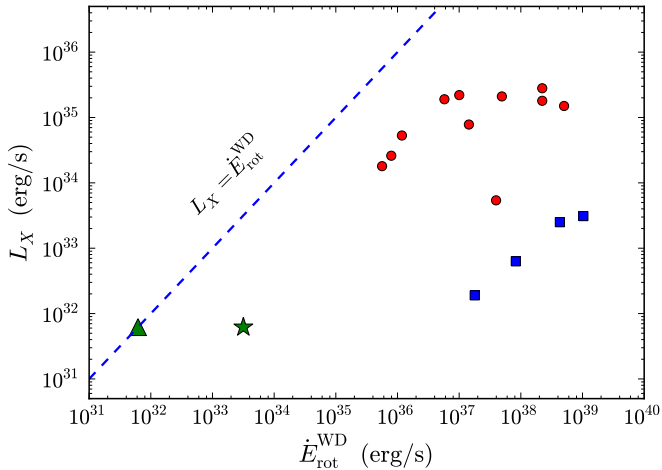}
\caption{X-ray luminosity $L_X$ versus the loss of rotational energy $\dot{E}_{\rm rot}$ describing SGRs and AXPs by rotation powered white dwarfs. The green star and the green triangle correspond to SGR 0418+5729 using respectively the upper and the lower limit of $\dot{P}$ given by Eq.~(\ref{eq:Pdotnew}). The blue squares are the only four sources that satisfy $L_X<\dot{E}_{\rm rot}$ when described as neutron stars (see Fig.~\ref{fig:LxvsEdotNS} for details).}
\label{fig:LxvsEdotWD}
\end{figure}

\begin{figure}[h]
\includegraphics[width=\hsize,clip]{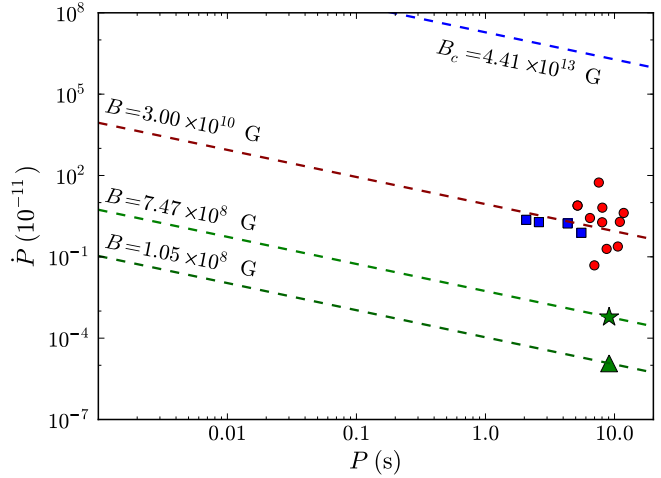}
\caption{$\dot{P}$-$P$ diagram for all known SGRs and AXPs. The curves of constant magnetic field for white dwarfs given by Eq.~(\ref{eq:BmaxWD}) are shown. The blue dashed line corresponds to the critical magnetic field $B_c=m^2_e c^3/(e \hbar)$. The green star and the green triangle correspond to SGR 0418+5729 using respectively the upper and the lower limit of $\dot{P}$ given by Eq.~(\ref{eq:Pdotnew}). The blue squares are the only four sources that satisfy $L_X<\dot{E}_{\rm rot}$ when described as rotation powered neutron stars (see Fig.~\ref{fig:LxvsEdotNS} for details).}
\label{fig:ppdotWD}
\end{figure}

\begin{figure}[h]
\includegraphics[width=\hsize,clip]{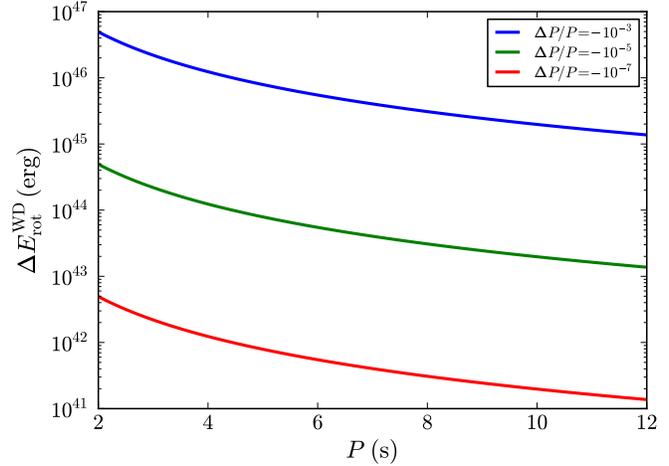}
\caption{Change in the rotational energy of the white dwarf $\Delta E^{\rm WD}_{\rm rot}$ given by Eq.~(\ref{eq:deltaEglitch}) as a function of the rotational period $P$ in seconds for selected fractional changes of period $\Delta P/P$.}
\label{fig:glitchesWD}
\end{figure}

Our aim in the following is to show that this model can be also applied to the other SGRs and AXPs. Their entire energetics is explained by the rotational energy loss of fast rotating magnetized white dwarfs: 1) the X-ray luminosity is well below the rotational energy loss of the white dwarf (see Fig.~\ref{fig:LxvsEdotWD}); 2) in all cases the large magnetic field is well below the critical field for vacuum polarization (see Fig.~\ref{fig:ppdotWD} and Table \ref{tab:properties}); 3) the energetics of all the bursts can be simply related to the change of rotational energy implied by the observed change of rotational period (see Fig.~\ref{fig:glitchesWD}, Sec.~\ref{sec:5} and Table \ref{tab:glitches}).

\section{SGRs and AXPs within the magnetar model}\label{sec:3}

\begin{figure}
\includegraphics[width=\hsize,clip]{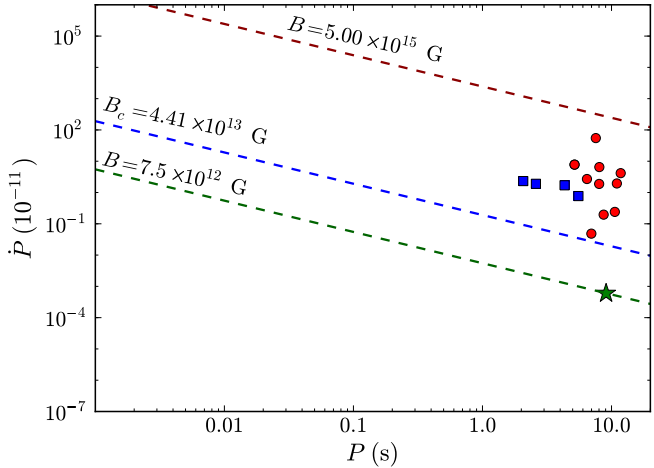}
\caption{$\dot{P}$-$P$ diagram for all known SGRs and AXPs. The curves of constant magnetic field for neutron stars given by Eq.~(\ref{eq:BmaxNS}) are shown. The blue dashed line corresponds to the critical magnetic field $B_c=m^2_e c^3/(e \hbar)$. The green star corresponds to SGR 0418+5729 using the upper limit of $\dot{P}$ given by Eq.~(\ref{eq:Pdotnew}). The blue squares are the only four sources that satisfy $L_X<\dot{E}_{\rm rot}$ when described as rotation powered neutron stars (see Fig.~\ref{fig:LxvsEdotNS} for details).}
\label{fig:ppdotNS}
\end{figure}

Let us turn to the alternative model commonly addressed as ``magnetar'' \citep[see e.g.][]{duncan92,thompson95} based on an ultramagnetized neutron star of $M = 1.4 M_\odot$ and $R=10$ km and then $I\approx 10^{45}$ g cm$^2$ as the source of SGRs and AXPs. The limit of the magnetic field obtained from Eq.~(\ref{eq:Bmax}) becomes
\begin{equation}\label{eq:BmaxNS}
B = 3.2\times 10^{19} \left(P \dot{P} \right)^{1/2} {\rm G}\, ,
\end{equation}
which is four orders of magnitude larger than the surface magnetic field within the fast rotating magnetized white dwarf model (see Fig.~\ref{fig:ppdotNS}).

\begin{figure}
\includegraphics[width=\hsize,clip]{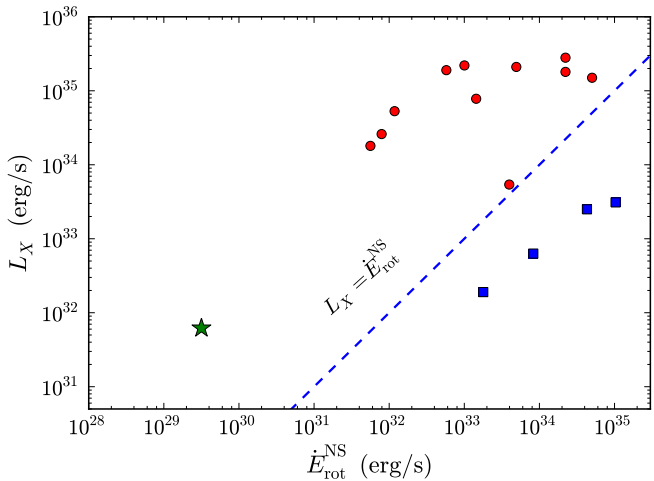}
\caption{X-ray luminosity $L_X$ versus the loss of rotational energy $\dot{E}_{\rm rot}$ describing SGRs and AXPs as neutron stars. The green star corresponds to SGR 0418+5729 using the upper limit of $\dot{P}$ given by Eq.~(\ref{eq:Pdotnew}). The blue squares are the only four sources with $L_X<\dot{E}_{\rm rot}$: 1E 1547.0-5408 with $P=2.07$ s and $\dot{P}=2.3\times 10^{-11}$; SGR 1627-41 with $P=2.59$ s and $\dot{P}=1.9\times 10^{-11}$; PSR J 1622-4950 with $P=4.33$ s and $\dot{P}=1.7\times 10^{-11}$; and XTE J1810--197 with $P=5.54$ s and $\dot{P}=7.7\times 10^{-12}$.}
\label{fig:LxvsEdotNS}
\end{figure}

There are innumerous papers dedicated to this model and for a review covering more than 250 references on the subject see \cite{mereghetti08}. The crucial point is that in this model there is no role of the rotational energy of the source: the X-ray luminosity is much bigger than the loss of rotational energy of the neutron star (see Fig.~\ref{fig:LxvsEdotNS}).

\begin{figure}[h]
\includegraphics[width=\hsize,clip]{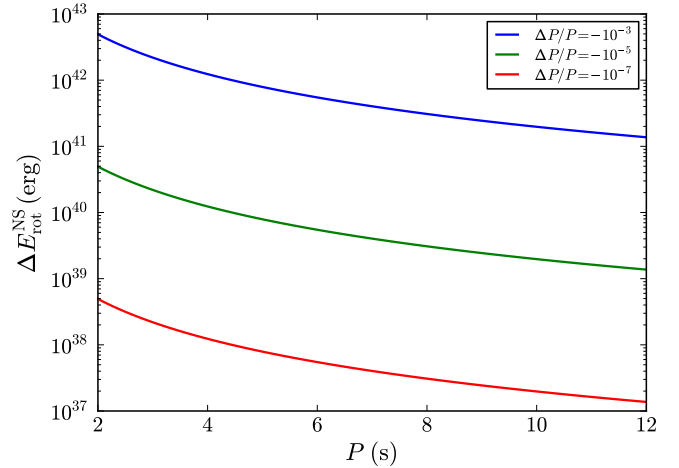}
\caption{Change in the rotational energy of the neutron star $\Delta E^{\rm NS}_{\rm rot}$ given by Eq.~(\ref{eq:deltaEglitchNS}) as a function of the rotational period $P$ in seconds for selected fractional changes of period $\Delta P/P$.}
\label{fig:glitchesNS}
\end{figure}

Paradoxically, although the bursts appear to be correlated to the presence of glitches in the rotational period, the corresponding increase of change of rotational energy of the neutron star 
\begin{equation}\label{eq:deltaEglitchNS}
\Delta E^{\rm NS}_{\rm rot} = -\frac{2 \pi^2 I}{P^2} \frac{\Delta P}{P} = -1.98\times 10^{46} \frac{\Delta P}{P^3}\quad {\rm erg}\, ,
\end{equation}
cannot explain the burst energetic $\sim (10^{44}$--$10^{47})$ erg. This is a clear major difference between the two models based respectively on neutron stars and white dwarfs (see Figs.~\ref{fig:glitchesWD} and \ref{fig:glitchesNS} for details).

In magnetars, the value of the rotational period and its first time derivative are only used to establish an upper limit to the magnetic field of the neutron star. In view of the smallness of the moment of inertia of a neutron star with respect to the moment of inertia of a white dwarf, the magnetic field reaches in many cases outstandingly large values $B >> B_c \sim 4.4\times 10^{13}$ G, from here the name magnetars (see Fig.~\ref{fig:ppdotNS}). The attempt has been proposed by \cite{duncan92} and \cite{thompson95} to assume a new energy source in physics and astrophysics: the magnetic energy in bulk. The role of thermonuclear energy has been well established by physics experiments on the ground as well as in astrophysics in the explanation of the energetics, life time, and build-up process of the nuclear elements in main sequence stars \citep[see e.g.][and references therein]{bethe68}; equally well established has been the role of rotational energy in pulsars \citep[see e.g.][and references therein]{hewish74,bell67}; similarly well established has been the role of gravitational energy in accretion process into neutron stars and black holes and binary X-ray sources \citep[see e.g.][and references therein]{giacconi02,giacconi78gral}. In the magnetars instead, it is introduced an alternative primary energy source not yet tested neither in the laboratory (the case of magnetic monopoles) nor in astrophysics: a primary energy source due to overcritical magnetic fields.

The mostly qualitative considerations in the magnetar model can be summarized, see e.g.~\cite{ng10}: in the twisted magnetosphere model of magnetars \citep{thompson02}, the observed X-ray luminosity of a magnetar is determined both by its surface temperature and by magnetospheric currents, the latter due to the twisted dipolar field structure. The surface temperature in turn is determined by the energy output from within the star due to magnetic field decay, as well as on the nature of the atmosphere and the stellar magnetic field strength. This surface thermal emission is resonantly scattered by the current particles, thus resulting in an overall spectrum similar to a Comptonized blackbody \citep[e.g.][]{lyutikov06,rea08,zane09}. In addition, the surface heating by return currents is believed to contribute substantially to $L_X$, at least at the same level as the thermal component induced from the interior field decay \citep{thompson02}. Magnetar outbursts in this picture occur with sudden increases in twist angle, consistent with the generic hardening of magnetar spectra during outbursts \citep[e.g.][]{kaspi03,woods04,israel07}.

It is worth to recall that magnetic field configurations corresponding to a dipole twisted field have been routinely adopted in rotating neutron stars \citep[see e.g.][]{cohen73}. Magnetic field annihilation and reconnection have been analogously adopted in solar physics \citep[see e.g.][]{parker57,sweet58} and also magnetic instabilities have been routinely studied in Tokamak \citep[see e.g.][]{coppi76}. These effects certainly occur in magnetized white dwarfs. What is important to stress here is that in none of these systems the magnetic field has been assumed to be the primary energy source of the phenomena, unlike in magnetars.

It is appropriate to recall just a few of the difficulties of the magnetar model in fitting observations, in addition to the main one of SGR 0418+5729 addressed in this article. In particular, e.g.: (1) as recalled by S.~Mereghetti 2008, ``up to now, attempts to estimate the magnetic field strength through the measurement of cyclotron resonance features, as successfully done for accreting pulsars, have been inconclusive''; (2) the prediction of the high-energy gamma ray emission expected in the magnetars has been found to be inconsistent with the recent observation of the Fermi satellite \citep[see e.g.][]{tong10,tong11}; (3) finally, it has been shown to be not viable the attempt to relate magnetars to the energy of the supernova remnants \citep[see e.g.][]{allen04,ferrario06,vink06,vink08} or to the formation of black holes (see e.g.~\cite{kasen10,woosley10}, see however e.g.~\cite{patnaude09}) and of Gamma Ray Bursts (see e.g.~\cite{levan06,castro08,stefanescu08,bernardini09}, see however e.g.~\cite{goldstein11,rea11}).

In Table \ref{tab:properties} we compare and contrast the parameters of selected SGRs and AXPs sources in the magnetar model and in the fast rotating highly magnetized white dwarf model: the larger radius of a white dwarf with respect to the radius of a neutron star of the same mass $M=1.4 M_\odot$, leads to the two models differing on the scale of mass density, moment of inertia, and rotational energy which imply a different scale for the surface magnetic fields, leading to a very different physical interpretation of the observations of SGRs and AXPs.

\section{Observations of massive fast rotating highly magnetized white dwarfs}\label{sec:4}

Some general considerations are appropriate. The white dwarf model appeals to standard and well tested aspects of physics and astrophysics. The observation of fast rotating white dwarfs with magnetic fields larger than $10^6$ G all the way up to $10^9$ G has been in the mean time solidly confirmed by observations \citep[see e.g][]{angel81,ferrario97,nalezyty04,ferrario05,terada08}. For a recent and extensive analysis of the magnetic field structure of highly magnetized white dwarfs see \cite{2009A&A...506.1341K} and for a catalog of them see \cite{2010yCat..35061341K} and also \cite{2010AIPC.1273...19K}.

A specific example is the highly magnetized white dwarf AE Aquarii. The rotational period of this fast rotating magnetized white dwarf obtained from the sinusoidal pulsed flux in soft X-rays $< 4$ keV  \citep[see e.g.][]{1991ApJ...370..330E,2006ApJ...646.1149C} has been established to be $P=33$ s and it is spinning down at a rate $\dot{P}=5.64\times 10^{-14}$. The mass of the white dwarf is $\sim M_\odot$ \citep{jager94} and the observed temperature is $kT \sim 0.5$ keV. In addition to the soft X-ray component, hard X-ray pulsations were observed with the Japanese satellite Suzaku in October-November 2005 and October 2006. The luminosity of AE Aquarii $\sim 10^{31}$ erg/s  accounts for the $0.09\%$ of the spin-down energy of the white dwarf \citep[see][for details]{terada08} and the infereed magnetic field of the source is $B \sim 10^8$ G \citep{ikhsanov08}. 

This white dwarf is one of the most powerful particle accelerators: there is at least one event of detected TeV emission from this source during its optical flaring activity monitored between 1988 and 1992 \citep[see e.g.][]{1992ApJ...401..325M,1993ICRC....1..338M,jager94,ikhsanov06,ikhsanov08,2011PhRvD..83b3002K}. In addition, it shows burst activity in X-rays \citep{terada08}. Although AE Aquarii is a binary system with orbital period $\sim 9.88$ hr \citep[see][e.g.]{jager94}, very likely the power due to accretion of matter is inhibited by the fast rotation of the white dwarf \citep[e.g.][]{2006ApJ...639..397I,terada08}. 

Many of the observed physical properties of this white dwarf are very similar to the recently discovered SGR 0418+5729, as we explicitly show in Table \ref{tab:WD-SGR}.
\begin{table}
\centering
\begin{tabular}{l c c }
\hline \hline
& {\footnotesize SGR 0418+5729} & {\footnotesize AE Aquarii}  \\
\hline
\vspace{0.2cm}
$P$ (s) & 9.08 & 33.08 \\
\vspace{0.2cm}
$\dot{P}$ ($10^{-14}$) & $< 0.6$ & $5.64$ \\
\vspace{0.2cm}
Age (Myr) & 24 & 9.4 \\
\vspace{0.2cm}
$L_X$ (erg/s) & $6.2\times 10^{31}$ & $\sim 10^{31}$ \\
\vspace{0.2cm}
$kT$ (keV) & 0.67 & 0.5 \\
\vspace{0.2cm}
$B$ (G) & $<7.45 \times 10^8$ & $\sim 10^8$ \\
\vspace{0.2cm}
Pulsed Fraction & 0.3 & $\sim 0.2$--$0.3$ \\
\hline
\end{tabular}
\caption{Comparison of the observational properties of SGR 0418+5729 and the white dwarf AE Aquarii. For SGR 0418+5729 $P$, $\dot{P}$, and $L_X$ have been taken from \cite{rea10}. The characteristic age is given by Age = $P/(2 \dot{P})$ and the surface magnetic field $B$ is given by Eq.~(\ref{eq:BmaxWD}). The pulsed fraction of SGR 0418+5729 is taken from \cite{2010MNRAS.405.1787E} and the one of the white dwarf AE Aquarii from \cite{1991ApJ...370..330E} and \cite{2006ApJ...646.1149C}.}
\label{tab:WD-SGR}
\end{table}

Although very fast, AE Aquarii is not the fastest white dwarf observed. The rotational period obtained from the pulsed X-ray emission of RXJ 0648.0-4418, the white dwarf in the binary system HD49798/RXJ 0648.0-4418, is $P=13.2$ s \citep{1997ApJ...474L..53I}. This white dwarf is one of the most massive white dwarfs with $M = 1.28 \pm 0.05 M_\odot$ \citep[see][for details]{2009Sci...325.1222M}. Other very massive and highly magnetized white dwarfs are: REJ 0317-853 with $M \sim 1.35 M_\odot$ and $B\sim (1.7$--$6.6)\times 10^8$ G \citep[see e.g.][]{1995MNRAS.277..971B,2010A&A...524A..36K}; PG 1658+441 with $M \sim 1.31 M_\odot$ and $B\sim 2.3\times 10^6$ G \citep[see e.g.][]{1983ApJ...264..262L,1992ApJ...394..603S}; and PG 1031+234 with the highest magnetic field $\sim 10^9$ G \citep[see e.g.][]{1986ApJ...309..218S,2009A&A...506.1341K}. It is interesting to note that the most highly magnetized white dwarfs are massive as well as isolated \citep[see e.g.][for details]{nalezyty04}.

\section{Rotational instability of white dwarfs}\label{sec:5}

\begin{figure}
\includegraphics[width=\hsize,clip]{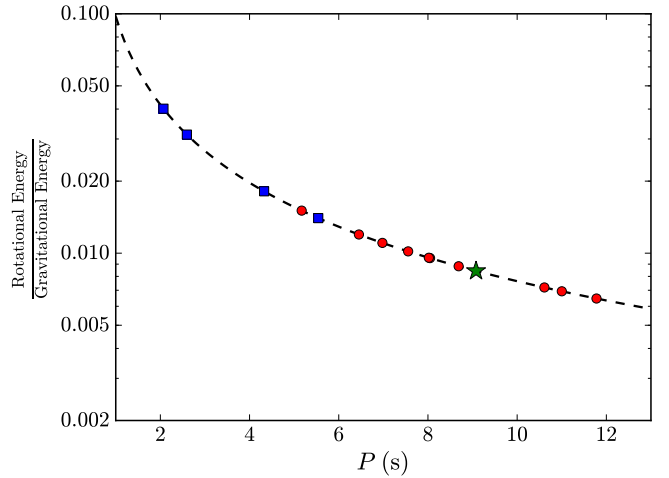}
\caption{Ratio between the rotational energy and the gravitational energy of a MacClaurin spheroid of $M=1.4M_\odot$ and $R=10^3$ km as a function of its rotational period $P$. The rotational period between 2 and 12 s appears to be very appropriate for fast rotating white dwarfs. Fast rotating neutron stars present much shorter period in the millisecond region. We show on the curve the position of all known SGRs and AXPs. The green star corresponds to SGR 0418+5729. The blue squares are the only four sources that satisfy $L_X<\dot{E}_{\rm rot}$ when described as rotation powered neutron stars (see Fig.~\ref{fig:LxvsEdotNS} for details).}
\label{fig:ToverW}
\end{figure}

In order to be stable against secular instability of the MacClaurin versus the Jacobi ellipsoid \citep{ferrari69}, the minimal period of a white dwarf with the parameters discussed here is $P_{\rm crit} \sim 0.94$ s. For $P\lesssim P_{\rm crit}$ we would expect very significant emission of gravitational waves due to the transition from the triaxial Jacobi ellipsoids to the axially symmetric MacClaurin ellipsoids. This is well in agreement and explains the observed long periods of SGRs and AXPs $\gtrsim 2$ s (see Fig.~\ref{fig:ToverW}). In the specific case of the source 1E 2259+586, assuming that the supernova remnant G109.1-1.0 and 1E 2259+586 are coeval, we obtain the initial rotational period of the white dwarf in the range $0.94$ s $<P_0<6.8$ s where, the lower limit, is given by the bifurcation point between MacClaurin spheroids and Jacobi ellipsoids \citep[see e.g.][]{ferrari69} and, the upper limit, is obtained for a constant value of $\dot{P}$. Describing today 1E 2259+586 by a MacClaurin spheroid, we obtain the ratio between the rotational energy and the gravitational energy $E_{\rm rot}/\left|E_{\rm grav}\right| \sim 0.011$ (see Fig.~\ref{fig:ToverW}), well below the secular instability $\sim 0.14$ and the dynamical instability $\sim 0.25$ \citep[see][for details]{chandrasekharbookb,shapirobook}.

The above considerations add interest in the recent theoretical analysis of white dwarfs taking into account nuclear, weak and electromagnetic interactions within a general relativistic treatment \citep{rueda10wd}. A specially relevant result has been recently obtained \citep{kuantay11a} by analyzing a white dwarf endowed with mass, angular momentum, and quadrupole moment within the Hartle-Thorne formalism \citep{1967ApJ...150.1005H,1968ApJ...153..807H}. The rotating white dwarfs have been studied for the new equation of state given by \cite{2011PhRvC..83d5805R} used for the construction of the non-rotating configurations by \cite{rueda10wd}. The critical rotational periods for the onset of the axisymmetric, the mass-shedding and the inverse $\beta$-decay instabilities have been studied in detail. The exact value of the critical period of a white dwarf depends upon the central density of the configuration; rotationally stable white dwarfs exist for rotational periods $P>P^{\rm WD}_{\rm min}\sim 0.3$ s. The shortest values for configurations supported by rotation with critical masses larger than the classical Chandrasekhar limit for non-rotating white dwarfs all the way up to $M_{\rm max}\sim 1.5 M_\odot$ \citep[see][for details]{kuantay11a}.

Consequently, also the fastest sources e.g. 1E 1547.0-5408 with $P=2.07$ s, SGR 1627-41 with $P=2.59$ s, and PSR J 1622-4950 with $P=4.33$ s, can be safely described as massive fast rotating white dwarfs as shown in Fig.~\ref{fig:LxvsEdotWD}.

\section{Glitches and outbursts in SGRs and AXPs}\label{sec:6}

The energetic of the observed bursts within the white dwarf model of SGRs and AXPs can be fully explained by the observed change of period $\Delta P<0$ (glitches). In the case of the famous event of 5th March 1979 in the SGR 0526-66 ($P=8.05$ s), a fractional change of period of the white dwarf $\Delta P/P \sim -10^{-4}$ (see Fig.~\ref{fig:glitchesWD}) would be sufficient to explain the energetics $\sim 3.6\times 10^{44}$ erg \citep{mereghetti08}. Unfortunately, such a change of period could not be observed at the time \citep[see e.g.][]{mazets79}, lacking the observations of the source prior to the event. Instead, in the case of the flares of 1E 2259+586 on June 2002 ($P=6.98$ s) and of 1E 1048.1-5937 ($P=6.45$ s) on March 2007, observational data are available. For 1E 2259+586, using the observed fractional change of period $\Delta P/P \sim -4 \times 10^{-6}$ \citep{woods04} (see also Fig.~\ref{fig:glitch1E2259}), we obtain within the white dwarf model a change of rotational energy $\left|\Delta E^{\rm WD}_{\rm rot} \right| \sim 1.7\times 10^{43}$ erg, to be compared with the measured energy released during the event $\sim 3\times 10^{41}$ erg. For the glitch on the 26th March 2007 in 1E 1048.1-5937 with observed $\Delta P/P \sim -1.63 \times 10^{-5}$, we obtain $\left|\Delta E^{\rm WD}_{\rm rot} \right| \sim 7.73 \times 10^{43}$ erg which is strikingly in agreement (and safely superior) with the observed energy released in the event $4.3\times 10^{42}$ erg \citep[see e.g.][]{2009ApJ...702..614D}. In the case of super giant flares, there is no clear observational evidence of their association to glitches. However, changes in the moment of inertia of the white dwarf originating fractional changes of period of order $\Delta P/P \sim -(10^{-5}-10^{-3})$ (see Fig.~\ref{fig:glitchesWD}) could explain their large energetics ranging from $10^{44}$ erg all the way up to $10^{47}$ erg \citep[see e.g.][]{mereghetti08}. For the giant flare of SGR 1806-20 on 27th December 2004 \citep[see e.g.][]{2004GCN..2920....1B,2005Natur.434.1098H} with observed energy $\sim 10^{46}$ erg there is a gap of timing data of the source between October 2004 and March 2005 \citep[see][]{2005ApJ...628..938M,2005A&A...440L..63T}. The observed rotational period of SGR 1806-20 after March 2005 is not consistent with the expected rotational period obtained from the spin-down rate $\dot{P}=5.5\times 10^{-10}$; instead, this is consistent with $\dot{P}=1.8\times 10^{-10}$. The change of rotational period has been attributed to ``global reconfigurations of the neutron star magnetosphere'' \citep[see e.g.][]{2005A&A...440L..63T}. Within the white dwarf model, such a burst activity is consistent with a glitch with fractional change of period $\sim -3\times 10^{-3}$. All the above discussion is summarized in Table \ref{tab:glitches} and Figs.~\ref{fig:glitch1E2259} and \ref{fig:glitchesWD}. 

In all the above cases the gain of rotational energy in the glitch is much larger than the energy observed in the flaring activities following the glitches. This means that there is ample room to explain these glitch-outburst events in a large range of recovery fractions $Q$. It appears to be appropriate to systematically monitor the $Q$ factors for all the glitches in SGRs and AXPs.
\begin{table*}
\centering
{\footnotesize
\begin{tabular}{l c c c c}
\hline \hline
& SGR 0526-66 & 1E 2259+586 & 1E 1048.1-5937 & SGR 1806-20 \\
\hline
Date & March 1979 & June 2002 & March 2007 & December 2004\\
Observed Energy (erg) & $3.6\times 10^{44}$ & $3\times 10^{41}$ & $4.2\times 10^{42}$ & $\sim 10^{46}$ \\
$|\Delta P|/P$ & $1.2\times 10^{-4}$ (predicted) & $4.24\times 10^{-6}$ (observed) & $1.63\times 10^{-5}$ (observed) & $3\times 10^{-3}$ (predicted)\\
Predicted Energy (erg) & $3.6\times 10^{44}$ & $1.7\times 10^{43}$ & $7.7\times 10^{43}$ & $\sim 10^{46}$ \\
\hline
\end{tabular}
}
\caption{Glitches and Outbursts of some SGRs and AXPs within the white dwarf model. The predicted values of $|\Delta P|/P$ are calculated with Eq.~(\ref{eq:deltaEglitch}) assuming $\left|\Delta E^{\rm WD}_{\rm rot} \right|$ equals the observed energy of the burst event. The predicted values of the energy released in the burst event is calculated with Eq.~(\ref{eq:deltaEglitch}) using the observed fractional change of rotational period $|\Delta P|/P$.}\label{tab:glitches}
\end{table*}

It is interesting that PSR J1846-0258, $P=0.3$ s, experienced in June 2006 a radiative event with estimated isotropic energy $\sim (3.8$--$4.8)\times 10^{41}$ erg \citep{2008ApJ...678L..43K}. Assuming that such an event was triggered by a glitch in the neutron star one obtains an associated fractional change of period $\Delta P/P \sim -(1.73$--$2.2)\times 10^{-6}$, as given by Eq.~(\ref{eq:deltaEglitchNS}). Indeed, as shown by \cite{2009A&A...501.1031K}, the outburst emission was accompanied by a large glitch $\Delta P/P \sim -(2.0$--$4.4)\times 10^{-6}$ in perfect agreement with the theoretical prediction given by the loss of rotational power after the spin-up of the neutron star without advocate any magnetar phenomena. This fact reinforces the idea that PSR J1846-0258 is not a magnetar but an ordinary rotationally powered neutron star, also in line with the recent suggestions by \cite{2009A&A...501.1031K} and \cite{rea10}.

\section{Magnetosphere emission from white dwarfs}\label{sec:7}

We return now to the structure of the magnetosphere of the white dwarf model for SGRs and AXPs. In order to have an agreement between the observed X-ray luminosity and the X-ray spectral distribution, it is necessary that only a part of the surface of the white dwarf has to be X-ray emitter.

We can define the dimensionless filling factor
\begin{equation}\label{eq:Rfactor}
{\cal R}=\frac{L_X}{4\pi R^2 \sigma T^4}\, ,
\end{equation}
where $\sigma$ is the Stefan-Boltzmann constant and $T$ the temperature of the source. This factor gives an estimate of the effective area of X-ray emission and consequently information about the structure of the magnetic field from the surface of the object. It is interesting that this factor for the white dwarf is in the range $10^{-6}$--$10^{-5}$ (see Table \ref{tab:properties}), quite similar to the one of the Sun ${\cal R}_\odot = L^X_\odot/(4 \pi R^2_\odot \sigma T^4_\odot) \approx (7.03\times 10^{-8}$--$1.2\times 10^{-6})$ in the minimum $L^X_\odot=2.7\times 10^{26}$ erg/s and in the maximum $L^X_\odot=4.7\times 10^{27}$ erg/s of solar activity respectively \citep[see e.g.][]{peres2000,judge03}. This should be expected by the general argument of the conservation of flux in the transition from a highly magnetized main sequence star to a white dwarf. The magnetic field of the order of $\sim 10^9$ G on the surface of these white dwarfs must clearly have a filamentary structure in the range ${\cal R}\sim 10^{-6}$--$10^{-5}$.

In the specific case of SGR 0418+572 such an ${\cal R}$ factor is $\sim 10^{-9}$, which is of the same order as the one of the white dwarf AE Aquarii, as can be seen from Table \ref{tab:WD-SGR} by comparing the values of $L_X$ and $K T$, which are the quantities involved in Eq.~(\ref{eq:Rfactor}).

At times the presence of an ${\cal R}$ factor has been interpreted as originating from a spot-like radial emission of the radiation from the surface of the white dwarf. If one were to assume that the radiation occurs radially beamed and occurring just from the surface either of the neutron star or the white dwarf, a spot radiation would lead to a pulsed fraction of the emission flux $ \sqrt{1/n \sum_{i=1}^n (y_i-\bar{y})^2}/\bar{y} \sim 1$ where $n$ is the number of phase bins per cycle, $y_i$ is the number of counts in the $i$th phase bin and $\bar{y}$ is the mean number of counts in the cycle \citep[see e.g.][for details about this definition]{2010MNRAS.405.1787E}. This problem, which seems to be in contradiction with the observations of pulsed fractions $<1$ in SGRs and AXPs \citep[see e.g.][]{2010MNRAS.405.1787E}, would be equally severe both for neutron stars and white dwarfs (see e.g.~Table \ref{tab:WD-SGR}).

It is appropriate to recall that all the SGRs and AXPs within a rotating white dwarf model have magnetic fields in the range $10^8\,{\rm G}\lesssim B \lesssim 10^{11}$ G (see Table \ref{tab:properties}). It is quite natural to assume that the X-ray emission be linked to the presence of the magnetic field. It is worth to note that the modeling of the physics and the geometrical structure of the magnetic field and of the magnetospheres is a most active field field of current research. As shown by \cite{romani2010}, the morphology of the pulses as well as of the light curves strongly depend on many model parameters, e.g. special and general relativistic effects, the viewing angle, the magnetic moment-spin axis angle, the spin axis-line of sight angle, the specific location of the emission zone, and the adopted magnetospheric model including possible corrections due to deviations from a pure dipolar structure.

From the broad sinusoidal pulsed flux of SGRs/AXPs \citep[see e.g.][]{mereghetti08}, we know that the pulsed fraction is less than one and that the luminosity differs remarkably from a spiky one. We find then natural to assume that the emission comes from an area covering the white dwarf surface with a very marked filamentary structure. Similar considerations for neutron stars magnetospheres have been purported e.g. by \cite{michel81}; \cite{michel83} giving evidence of magnetospheric activity from the pole all the way up to the equator; see also the most interesting case of the pair production activities in the magnetosphere of a rotating white dwarf considered for the transient radio source GCRT J1745--3009 by \cite{zhang05}. Moreover, such structures are regularly observed in the Sun and in the Earth Aurora. Explicit sinusoidal pulsed flux in soft X-rays ($<$ 4 keV) have been observed in AE Aquarii \citep[see e.g.][]{1991ApJ...370..330E,2006ApJ...646.1149C}; and see also Fig.~6 in \cite{mereghettiwd2011} for similar sinusoidal pulsed emission of the white dwarf RXJ 0648.0-4418 with rotational period $P=13.2$ s. For all the above sources, a filamentary structure of the magnetic field is clearly expected.

We do not discuss here the issue of the spectral features within the white dwarf model. The aim of this article is just to point out that all these problems can be address with merit starting from the rotational energy of a rotating white dwarf rather than the magnetic energy of a magnetar. The spectrum of the persistent emission of SGRs and AXPs for energies $<10$ keV is well fitted either by the superposition of a blackbody and a high energy tail or by a single blackbody or a double blackbody \citep[see e.g.][]{mereghetti08}. Such a spectral feature is clearly already evidenced for rotating white dwarfs; following the work of \cite{terada08}: in addition to the thermal modulation in the softer X-ray band, spiky pulsations like the ones of pulsars have been observed by the Suzaku satellite in the hard X-ray band of over 4 keV in the white dwarf AE Aquarii. The X-ray spectrum requires an additional hard X-ray component on the well-known thermal emissions with temperatures of 0.5 and 2.9 keV. Combined with results from timing analyses, spectral shapes and flux, it was there concluded that the hard X-ray pulsations should have a non-thermal origin, for example, possible Synchrotron emission with sub MeV electrons. The claim of the first discovery of a white dwarf equivalent to a neutron star pulsar was there made. In view of the possible evidence of very high energy emission in the TeV region observed during the optical flares of AE Aquarii \citep[see e.g.][and references therein]{jager94,ikhsanov06,ikhsanov08,terada08,2008AdSpR..41..512T,2011PhRvD..83b3002K}, it would be important to have observations by INTEGRAL and Fermi of rotating magnetized white dwarf in the 20-200 keV band in order to establish further analogies between fast rotating highly magnetized white dwarfs and magnetar candidates.

More specifically, for the source SGR 0418+5729 and its interpretation as a white dwarf, a crucial result has been recently obtained by \cite{durant2011}. We first recall the observed range of temperatures of massive isolated white dwarfs $1.14\times 10^4\,{\rm K}\leq T \leq 5.52\times 10^4$ K; see Table 1 in \citep{ferrario05b}. From the broad band Hubble Space Telescope imaging of the field of SGR 0418+5729, the upper limits of the black body surface temperature, $T<3.14\times 10^4$ K and $T<1.18\times 10^4$ K in the F110W and F606W filters, can be established for a radius $R=10^8$ cm. In this respect is also worth to recall the optical observations of AXP 4U0142+61 of \cite{hulleman2000}. The photometric results of the field of 4U0142+61 at the 60-inch telescope on Palomar Mountain are in agreement with a $1.3 M_\odot$ white dwarf with a surface temperature $\sim 4\times 10^5$ K \citep[see][for details]{hulleman2000}. These results are therefore fully consistent with the SGR/AXP white dwarf model, and follow-on missions of Hubble and VLT are strongly recommended.
 
\section{The connection with supernova remnants}\label{sec:8}

We would like to address the special issue of the supernova remnants energetics and their association with SGRs and AXPs. A firm association between SGRs/AXPs and supernovae have been purported by \cite{gaensler01} in the cases 1E 1841--045 (SNR G27.4+0.0, Kes 73), AX J1845.0--0258 (SNR G29.6+0.1), and 1E 2259+586 (SNR G109.1--1.0, CTB 109). See also \cite{2007ApJ...667.1111G} for the possible association 1E 1547.0-5408 (SNR G327.24-0.13). What is of interest for us here is the special issue of the energetics of the supernova remnant and the present of an SGR or an AXP.

Paczynski, in the case of AXP 1E 2259+586, attempted to explain the supernova remnant by assuming a merger of a binary system of ordinary white dwarf of mass $\sim (0.7$--$1) M_\odot$ based on models by \cite{iben84} and \cite{paczynski85} leading both to the formation of a fast rotating white dwarf and to the supernova remnant. Recent simulations of white dwarf-white dwarf mergers \citep[see e.g.][]{pakmor10} point that mergers of ($0.8$--$0.9 M_\odot$) produce supernova events generally not very efficient energetically, well below the observed explosion energy $\sim 7.4\times 10^{50}$ erg of the supernova remnant G109.1-1.0 associated to 1E 2259+586 \citep[see e.g.][]{sasaki04}.

In the intervening years much more has been understood on the process of gravitational collapse and on the composition of the material surrounding neutron stars and black holes both from pulsar observations and Gamma Ray Bursts. Fascinating evidence for the presence of planets around pulsars in supernova remnants has been established \citep[see e.g.][]{konacki99,hansen02,konacki03}. Similarly, the presence of many body process of gravitational collapse has been evidenced for Gamma Ray Bursts \citep[see e.g.][]{ruffinimg12}. 

In view of the above, we advance the possible scenario in which the SGRs/AXPs and the supernova remnant originate from a very close binary system composed of a white dwarf and a companion late evolved star, close to the process of gravitational collapse. The collapse of the companion star, either to a neutron star or to a black hole, leads to mass loss which can unbind the original binary system. Three possible cases can occur \citep[see e.g.][]{ruffini73}: if the loss of mass in the supernova explosion is $M_{\rm loss}< M/2$, being $M$ the total mass of the binary, the system holds bound; 2) if $M_{\rm loss}\sim M/2$ then the system becomes unbound and the white dwarf is expelled at nearly orbital motion velocity; and 3) if $M_{\rm loss}>> M/2$ the white dwarf is kicked out with very high runaway velocities. Only in the first case the object will lie at the center of the supernova remnant. For a review on the evolution of binary systems see \cite{2004Sci...304..547S} and for a detailed treatment of the problem of runaway velocities from supernova explosions see \cite{1996A&A...315..432T,1998A&A...330.1047T}. 

The white dwarf in this picture does not participate either to the gravitational collapse nor to the formation of the supernova remnant: it can have a period and a life time determine essentially by the prior evolution of the binary system. This explains the disagreement between the age of the supernova remnant and the characteristic age of the SGR/AXP when inferred by a neutron star model. In the case of large kick velocities the runaway white dwarf can collide with the surrounding material in the supernova remnant and very likely also with planets. Such collisions may well originate changes in the moment of inertia of the white dwarf, consequently in its rotational period, leading to glitches and burst activity.

In the above context it is appropriate to recall the pioneering work of \cite{1996ApJ...463..305K} on explaining the super-Eddington luminosities in the flaring episodes of SGRs and AXPs as originating in accretion process of planetary fragments, in particular, the important role of magnetic confinement of an $e^+e^-$ pair plasma. The model explains the observed self-absorbed thermal spectrum of flares and their nearly independence on their luminosity. \cite{1996ApJ...463..305K} has shown that the infall of planetary fragments may lead to a continuous injection of energy to the magnetosphere which leads to magnetic confinement of the source if the magnetic field satisfies
\begin{equation}
B>\sqrt{\frac{2L}{cR^2}} = 2.6\times 10^7 \sqrt{\frac{L_{41}}{R^2_8}}\quad {\rm G}\, ,
\end{equation}
where $L_{41}$ is the luminosity in units of $10^{41}$ erg/s and $R_8$ is the radius of the source in units of $10^8$ cm.

In the case when the radiation is not being continuously resupplied, but it is initially contained within the volume $\sim 4\pi R^3/3$, the minimum magnetic field for confinement is given by
\begin{equation}
B> \sqrt{\frac{6L \tau}{R^3}} = 2.45\times 10^8 \sqrt{\frac{L_{41} \tau_{0.1}}{R^3_8}}\quad {\rm G}\, ,
\end{equation}
where $\tau_{0.1}$ is the time $\tau$ during which the source is radiating at a luminosity $L$, in units of 0.1 s. The fiducial values for $L$ and for $\tau$ has been chosen here to be typical of the bursting activity of SGRs/AXPs \cite[see e.g.][]{mereghetti08}. The above two bounds for the magnetic field are indeed in line with the surface magnetic fields obtained in this paper; see Fig.~\ref{fig:ppdotWD} for details. Thus, the super-Eddington luminosities observed in the outbursts can be well explained within the white dwarf model and there is no need of introducing the huge magnetic fields of the magnetar model \citep[][]{paczynski92,thompson95}.

\section{On the fiducial neutron star and white dwarf parameters in light of recent theoretical progress}\label{sec:9}

Before concluding, we would like to introduce a word of caution on the fiducial values adopted both for the neutron star and the white dwarf in the above Sections. In the intervening years much more have been learned on the equation of state and on a more complex description of the structure parameters of both white dwarfs and neutron stars.

The equations of equilibrium of neutron stars, traditionally based on the Tolman-Oppenheimer-Volkoff equations, have been superseded by an alternative formulation based on the general relativistic Thomas-Fermi conditions of equilibrium within the Einstein-Maxwell equations \cite{ruedaNPA}. Correspondingly, the above values of $\sqrt{I/R^6}$ in Eq.~(\ref{eq:Bmax}) estimated int he fiducial parameters, leading to Eq.~(\ref{eq:BmaxNS}), can acquire in fact values in the range $0.44 \lesssim \sqrt{I/R^6}/\sqrt{I_f/R_f^6} \lesssim 0.56$, where the subscript `f' stands for fiducial parameter. This range corresponds to the range of masses $0.5 \lesssim M/M_\odot \lesssim 2.6$ \citep{belvedere2011}. Correspondingly, the magnetic field is in the range $0.44 \lesssim B/B^{\rm NS}_f \lesssim 0.56$, where $B^{\rm NS}_f$ is given by Eq.~(\ref{eq:BmaxNS}).

Similar considerations apply for the white dwarf case. General relativistic white dwarfs taking into account nuclear, weak and electromagnetic interactions have been recently constructed \citep{rueda10wd} following the new equation of state for compressed nuclear matter given by \cite{2011PhRvC..83d5805R}. The case of rotating white dwarfs in general relativity has been studied by \cite{kuantay11a}. It has been found that white dwarfs can be as fast as $P^{\rm WD}_{\rm min} \sim 0.3$ s and as massive as $M_{\rm max}\sim 1.5 M_\odot$; see Sec.~\ref{sec:5} for details. For example, a white dwarf of $M = 1.44 M_\odot$ rotating with period $P = 3.2$ s, will have an equatorial radius $R_{\rm eq} \sim 3604$ km, polar radius $R_p \sim 2664$ km, and moment of inertia $I \sim 2.9 \times 10^{49}$ g cm$^2$. In this case we will have $\sqrt{I/R^6}/\sqrt{I_f/R_f^6} \sim 0.01$ and therefore $B/B^{\rm WD}_f \sim 0.01$ where $B^{\rm WD}_f$ is given by Eq.~(\ref{eq:BmaxWD}).

This issue is particularly relevant to the study of the four sources in Fig.~\ref{fig:LxvsEdotNS}. These sources can be definitely explained within a unified framework of rotating white dwarfs with all the other SGRs and AXPs. In view of the parameters recently obtained they may be also interpreted as regular neutron stars with a barely critical magnetic field. For these sources an option remain open for their interpretation as white dwarfs or neutron stars. A more refined analysis will clarify the correctness of the two possible interpretations both, in any case, alternative to the magnetar model.

\section{Conclusions and remarks}\label{sec:10}

\begin{figure}
\includegraphics[width=\hsize,clip]{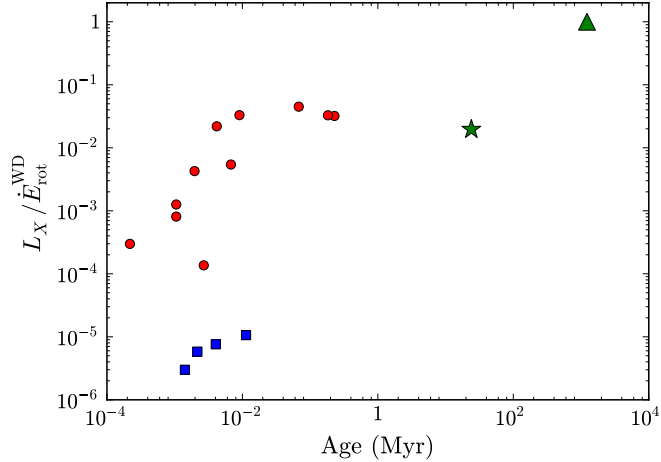}
\caption{Ratio between the observed X-ray luminosity $L_X$ and the loss of rotational energy $\dot{E}_{\rm rot}$ describing SGRs and AXPs by rotation powered white dwarfs. The green star and the green triangle correspond to SGR 0418+5729 using respectively the upper and the lower limit of $\dot{P}$ given by Eq.~(\ref{eq:Pdotnew}). The blue squares are the only four sources that satisfy $L_X<\dot{E}_{\rm rot}$ when described as rotation powered neutron stars (see Fig.~\ref{fig:LxvsEdotNS} for details).}
\label{fig:LxoverEdot}
\end{figure}

The recent observations of the source SGR 0418+5729 cast a firm separatrix in comparing and contrasting the two models for SGRs and AXPs based respectively on an ultramagnetized neutron star and on a white dwarf. The limit on the magnetic field derived in the case of a neutron star $B=7.5\times 10^{12}$ G makes it not viable as an explanation based on the magnetar model both from a global energetic point of view and from the undercritical value of the
magnetic field. In the white dwarf model, the picture is fully consistent. It is interesting that the rotational
energy loss appears to approach the value of the observed X-ray luminosity with time (see Fig.~\ref{fig:LxoverEdot}) as the magnetospheric activity settles down.

The description of SGR 0418+5729 as a white dwarf predicts the lower limit of the spin-down rate $\dot{P}$ given by Eq.~(\ref{eq:Pdotnew}), the surface magnetic field field is, accordingly to Eq.~(\ref{eq:BmaxWD}), constrained by $1.05\times 10^8\,\,{\rm G} < B_{\rm SGR 0418+5729} < 7.47\times 10^8\,\,{\rm G}$ (see Fig.~\ref{fig:ppdotWD}). The campaign of observations launched by the Fermi and Agile satellites will address soon this issue and settle in the near future this theoretical prediction.

The characteristic changes of period $\Delta P/P \sim -(10^{-7}$--$10^{-3}$) and the relating bursting activity $\sim (10^{41}$--$10^{46}$) erg in SGRs and AXPs can be well explained in term of the rotational energy released after the glitch of the white dwarf. It is also appropriate to recall that fractional changes, on scales $|\Delta P|/P \lesssim 10^{-6}$ are also observed in pulsars and routinely expressed in terms of the release of rotational energy of the neutron star, without appealing to any magnetars phenomena; e.g.~the glitch/outburst activity experienced in June 2006 by PSR J1846-0258 (see Sec.~\ref{sec:6}). 

In the magnetar model the dipole field is invoked to explain the period and the slowing down of the star leading to enormous magnetic fields $\sim 10^{14}$--$10^{15}$ G, see e.g.~Fig.~\ref{fig:ppdotNS}. The steady emission as well as the transient activity needs an additional explanation as due to the decay of strong multipolar magnetic fields \citep[see e.g.][and references therein]{tong11}. In the case of a model based on quark stars, a second component represented by an accretion disk around the star is also required to explain the energetics, without appealing to ultra-strong magnetic fields \citep[][]{xu06,tong11}. In the case of the model based on a rotating magnetized white dwarf, we show that the occurrence of the glitch, the associated sudden shortening of the period, as well as the corresponding gain of rotational energy, can be explained by the release of gravitational energy associated to a sudden contraction and decrease of the moment of inertia of the white dwarfs, consistent with the conservation of the angular momentum. The energetics of the steady emission as well as the one of the outbursts following the glitch can be simply explained in term of the loss of the rotational energy, in view of the moment of inertia of the white dwarfs, much larger than the one of neutron stars or quark stars, see Eqs.~(\ref{eq:DeltaEg}) and (\ref{eq:deltaEglitch}).

The observation of massive fast rotating highly magnetized white dwarfs by dedicated missions as the one leadered by the X-ray Japanese satellite Suzaku \citep[see e.g.][]{terada08} has led to the confirmation of the existence of white dwarfs sharing common properties with neutron star pulsars, hence their name white dwarf pulsars. The theoretical interpretation of the high-energy emission from white dwarf pulsars will certainly help to the understanding of the SGR and AXP phenomena \citep[see e.g.][]{2011PhRvD..83b3002K}.

We have given evidence that all SGRs and AXPs can be interpreted as rotating white dwarfs providing that the rotational period satisfies $P>P^{\rm WD}_{\rm min} \sim 0.3$ s.

Concerning the rotational period of SGRs and AXPs, it becomes interesting to confront our general relativistic results on uniformly rotating white dwarfs \citep{kuantay11a} with the interesting work of \cite{1968ApJ...151.1089O} on differentially rotating Newtonian white dwarfs. 

Regarding magnetized white dwarfs, the coupling between rotation and Rayleigh-Taylor instabilities arising from chemical separation upon crystallization may have an important role in the building of the magnetic field of
the white dwarf \cite{garcia2011}.

We encourage observational campaigns from space and ground for gaining understanding in the most fundamental issue of relativistic astrophysics: the identification of the SGRs/AXPs energy source.

We are grateful to the referee for a clear formulation of a number of fundamental issues which we have addressed and solved. We also thank for the careful reading of the manuscript and the many positive suggestions. We acknowledge Aldo Treves for crucial discussions. We also thank for interesting discussions on this subject to E.~Garc\'ia-Berro, Jorge Horvath, Jeremiah Ostriker, Jos\'e Pacheco, Yukikatsu Terada and Vladimir Usov, as well as to all the participants of the IRAP Erasmus Mundus Ph.~D. School ``From Nuclei to White Dwarfs to Neutron Stars'' held at Les Houches, France, April 2011. Manuel Malheiro acknowledges the hospitality and support of ICRANet, and the Brazilian agency FAPESP (fellowship BPE 2010/0558-1 in the thematic project 2007/03633-3) for the financial support.

\emph{Note added after submission}: We stress here the most recent observations of PSR J1841--0500 with rotation period $P=0.9$ s. This pulsar is located at only $4'$ from the AXP 1E 1841--045\footnote{The properties of 1E 1841--045 can be found in the fourth column of the lower half of Table \ref{tab:properties}.}, associated to the supernova remnant Kes 73 \citep[see][for details]{2011arXiv1111.5870C}. Such a discovery represents a clear observational support for the prediction of the binary scenario we introduced in Sec.~\ref{sec:8}, leading to an SGR/AXp, a supernova remnant and an additional neutron star or black hole. Deep searches for radio pulsations in the vicinities of the other sources AX J1845.0--0258, associated to SNR G29.6+0.1, 1E 2259+586, associated to SNR G109.1--1.0 (CTB 109), and 1E 1547.0-5408 associated to SNR G327.24-0.13, are highly recommended.

\begin{table*}
\centering
\begin{tabular}{l c c c c}
\hline \hline
& SGR 1806-20 & SGR 0526-66 & SGR 1900+14 & SGR 0418+5729 \\
\hline
\vspace{0.2cm}
$P$ (s) & 7.56 & 8.05 & 5.17 & 9.08\\
\vspace{0.2cm}
$\dot{P} (10^{-11})$ & 54.9 & 6.5 & 7.78 & $<6.0 \times 10^{-4}$\\
\vspace{0.2cm}
Age (kyr) & 2.22 & 1.97 & 1.05 & $24.0\times 10^3$\\
\vspace{0.2cm}
$L_X (10^{35}$ erg/s) & 1.50 & 2.1 & 1.8 & $6.2 \times 10^{-4}$\\
\vspace{0.2cm}
$kT$ (kev) & 0.65 & 0.53 & 0.43 & 0.67 \\
\vspace{0.2cm}
$\dot{E}^{\rm WD}_{\rm rot} (10^{37}$ erg/s) & 50.24 & 4.92 & 22.24 & $3.2 \times 10^{-4}$ \\
\vspace{0.2cm}
$B_{\rm WD} (10^9$ G) & 206.10 & 73.18 & 64.16 & 0.75\\
\vspace{0.2cm}
${\cal R}_{\rm WD} (10^{-5})$ & 0.65 & 2.06 & 4.07 & $2.4\times 10^{-4}$\\
\vspace{0.2cm}
$\dot{E}^{\rm NS}_{\rm rot} (10^{35}$ erg/s) & 0.502 & 0.05 & 0.22 & $3.2 \times 10^{-6}$\\
\vspace{0.2cm}
$B_{\rm NS} (10^{14}$ G) & 20.61 & 7.32 & 6.42 & 0.075\\
${\cal R}_{\rm NS}$ & 0.065 & 0.21 & 0.41 & $2.4\times 10^{-5}$\\
\hline \hline
& 1E 1547-54 & 1E 1048-59  & 1E 1841-045 & 1E 2259+586 \\
\hline
\vspace{0.2cm}
$P$ (s) & 2.07 & 6.45 & 11.78 & 6.98\\
\vspace{0.2cm}
$\dot{P} (10^{-11})$ & 2.32 & 2.70 & 4.15 & 0.048\\
\vspace{0.2cm}
Age (kyr) & 1.42 & 3.79 & 4.50 & 228.74\\
\vspace{0.2cm}
$L_X (10^{35}$ erg/s) & 0.031 & 0.054 & 2.2 & 0.19\\
\vspace{0.2cm}
$kT$ (kev) & 0.43 & 0.62 & 0.38 & 0.41\\
\vspace{0.2cm}
$\dot{E}^{\rm WD}_{\rm rot} (10^{37}$ erg/s) & 103.29 & 3.97  & 1.01 & 0.056 \\
\vspace{0.2cm}
$B_{\rm WD} (10^9$ G) & 22.17 & 42.22 & 70.71 &  5.88\\
\vspace{0.2cm}
${\cal R}_{\rm WD} (10^{-5})$ & 0.07 & 0.028 & 8.16 & 0.49\\
\vspace{0.2cm}
$\dot{E}^{\rm NS}_{\rm rot} (10^{35}$ erg/s) & 1.03 & 0.040 & 0.010 & $5.62\times 10^{-4}$\\
\vspace{0.2cm}
$B_{\rm NS} (10^{14}$ G) & 2.22 & 4.22 & 7.07 &  0.59\\
${\cal R}_{\rm NS}$ & 0.007 &	0.0028 & 0.82 & 0.049 \\
\hline
\end{tabular}
\caption{SGRs and AXPs as white dwarfs and neutron stars. The rotational period $P$, the spin-down rate $\dot{P}$, the X-ray luminosity $L_X$ and the temperature $T$ have been taken from the McGill online catalog at www.physics.mcgill.ca/$\sim$pulsar/magnetar/main.html. The characteristic age is given by Age = $P/(2 \dot{P})$, the loss of rotational energy $\dot{E}_{\rm rot}$ is given by Eqs.~(\ref{eq:Edot}) and Eq.~(\ref{eq:EdotNS}) and the surface magnetic field is given by Eqs.~(\ref{eq:BmaxWD}) and (\ref{eq:BmaxNS}) for white dwarfs and neutron stars respectively. The filling factor ${\cal R}$ is given by Eq.~(\ref{eq:Rfactor}).}\label{tab:properties}
\end{table*}




\begin{thebibliography}{122}
\expandafter\ifx\csname natexlab\endcsname\relax\def\natexlab#1{#1}\fi

\bibitem[{{Allen} \& {Horvath}(2004)}]{allen04}
{Allen}, M.~P. \& {Horvath}, J.~E. 2004, \apj, 616, 346

\bibitem[{{Althaus} {et~al.}(2005){Althaus}, {Garc{\'{\i}}a-Berro}, {Isern}, \&
  {C{\'o}rsico}}]{2005A&A...441..689A}
{Althaus}, L.~G., {Garc{\'{\i}}a-Berro}, E., {Isern}, J., \& {C{\'o}rsico},
  A.~H. 2005, \aap, 441, 689

\bibitem[{{Althaus} {et~al.}(2007){Althaus}, {Garc{\'{\i}}a-Berro}, {Isern},
  {C{\'o}rsico}, \& {Rohrmann}}]{2007A&A...465..249A}
{Althaus}, L.~G., {Garc{\'{\i}}a-Berro}, E., {Isern}, J., {C{\'o}rsico}, A.~H.,
  \& {Rohrmann}, R.~D. 2007, \aap, 465, 249

\bibitem[{{Angel} {et~al.}(1981){Angel}, {Borra}, \& {Landstreet}}]{angel81}
{Angel}, J.~R.~P., {Borra}, E.~F., \& {Landstreet}, J.~D. 1981, \apjs, 45, 457

\bibitem[{{Barstow} {et~al.}(1995){Barstow}, {Jordan}, {O'Donoghue},
  {Burleigh}, {Napiwotzki}, \& {Harrop-Allin}}]{1995MNRAS.277..971B}
{Barstow}, M.~A., {Jordan}, S., {O'Donoghue}, D., {et~al.} 1995, \mnras, 277,
  971

\bibitem[{{Baym} \& {Pines}(1971)}]{1971AnPhy..66..816B}
{Baym}, G. \& {Pines}, D. 1971, Annals of Physics, 66, 816

\bibitem[{{Belvedere} {et~al.}(2011){Belvedere}, {Pugliese}, {Rueda}, {Ruffini}, \& {Xue}}]{belvedere2011}
{Belvedere}, R., {Pugliese}, D., {Rueda}, J.~A., {Ruffini}, R., \& {Xue}, S. 2011, submitted to Nucl.\ Phys.\ A.

\bibitem[{{Bell} \& {Hewish}(1967)}]{bell67}
{Bell}, S.~J. \& {Hewish}, A. 1967, Nature, 213, 1214

\bibitem[{{Bernardini} {et~al.}(2009){Bernardini}, {Israel}, {Dall'Osso},
  {Stella}, {Rea}, {Zane}, {Turolla}, {Perna}, {Falanga}, {Campana},
  {G{\"o}tz}, {Mereghetti}, \& {Tiengo}}]{bernardini09}
{Bernardini}, F., {Israel}, G.~L., {Dall'Osso}, S., {et~al.} 2009, \aap, 498,
  195

\bibitem[{{Bethe}(1968)}]{bethe68}
{Bethe}, H.~A. 1968, {Energy production in stars. Nobel lecture.}, ed. {Bethe,
  H.~A.}

\bibitem[{{Borkowski} {et~al.}(2004){Borkowski}, {Gotz}, {Mereghetti},
  {Mowlavi}, {Shaw}, \& {Turler}}]{2004GCN..2920....1B}
{Borkowski}, J., {Gotz}, D., {Mereghetti}, S., {et~al.} 2004, GRB Coordinates
  Network, 2920, 1

\bibitem[{{Boshkayev} {et~al.}(2011){Boshkayev}, {Rueda}, \&
  {Ruffini}}]{kuantay11a}
{Boshkayev}, K., {Rueda}, J.~A., \& {Ruffini}, R. 2011, in From Nuclei to White
  Dwarfs to Neutron Stars, ed. {A.~Mezzacappa \& R.~Ruffini}

\bibitem[{{Camilo} {et~al.}(2011){Camilo}, {Ransom}, {Chatterjee},
  {Johnston}, \& {Demorest}}]{2011arXiv1111.5870C}
{Camilo}, F., {Ransom}, S.~M., {Chatterjee}, S., {Johnston}, S.,  \& {Demorest}, P. 2011, arXiv:1111.5870

\bibitem[{{Castro-Tirado} {et~al.}(2008){Castro-Tirado}, {de Ugarte Postigo},
  {Gorosabel}, {Jel{\'{\i}}nek}, {Fatkhullin}, {Sokolov}, {Ferrero}, {Kann},
  {Klose}, {Sluse}, {Bremer}, {Winters}, {Nuernberger},
  {P{\'e}rez-Ram{\'{\i}}rez}, {Guerrero}, {French}, {Melady}, {Hanlon},
  {McBreen}, {Leventis}, {Markoff}, {Leon}, {Kraus}, {Aceituno}, {Cunniffe},
  {Kub{\'a}nek}, {V{\'{\i}}tek}, {Schulze}, {Wilson}, {Hudec}, {Durant},
  {Gonz{\'a}lez-P{\'e}rez}, {Shahbaz}, {Guziy}, {Pandey}, {Pavlenko}, {Sonbas},
  {Trushkin}, {Bursov}, {Nizhelskij}, {S{\'a}nchez-Fern{\'a}ndez}, \&
  {Sabau-Graziati}}]{castro08}
{Castro-Tirado}, A.~J., {de Ugarte Postigo}, A., {Gorosabel}, J., {et~al.}
  2008, Nature, 455, 506

\bibitem[{{Chandrasekhar}(1969)}]{chandrasekharbookb}
{Chandrasekhar}, S. 1969, {Ellipsoidal figures of equilibrium}, ed.
  {Chandrasekhar, S.}

\bibitem[{{Choi} \& {Dotani}(2006)}]{2006ApJ...646.1149C}
{Choi}, C.-S. \& {Dotani}, T. 2006, \apj, 646, 1149

\bibitem[{{Cohen} {et~al.}(1973){Cohen}, {Coppi}, \& {Treves}}]{cohen73}
{Cohen}, R.~H., {Coppi}, B., \& {Treves}, A. 1973, \apj, 179, 269

\bibitem[{{Coppi} {et~al.}(1976){Coppi}, {Pellat}, {Rosenbluth}, {Rutherford},
  \& {Galvao}}]{coppi76}
{Coppi}, B., {Pellat}, R., {Rosenbluth}, M., {Rutherford}, P., \& {Galvao}, R.
  1976, Soviet Journal of Plasma Physics, 2, 961

\bibitem[{{Corbet} {et~al.}(1995){Corbet}, {Smale}, {Ozaki}, {Koyama}, \&
  {Iwasawa}}]{1995ApJ...443..786C}
{Corbet}, R.~H.~D., {Smale}, A.~P., {Ozaki}, M., {Koyama}, K., \& {Iwasawa}, K.
  1995, \apj, 443, 786

\bibitem[{{Davies} {et~al.}(1990){Davies}, {Coe}, \& {Wood}}]{davies90}
{Davies}, S.~R., {Coe}, M.~J., \& {Wood}, K.~S. 1990, \mnras, 245, 268

\bibitem[{{de Jager} {et~al.}(1994){de Jager}, {Meintjes}, {O'Donoghue}, \&
  {Robinson}}]{jager94}
{de Jager}, O.~C., {Meintjes}, P.~J., {O'Donoghue}, D., \& {Robinson}, E.~L.
  1994, \mnras, 267, 577

\bibitem[{{Dib} {et~al.}(2009){Dib}, {Kaspi}, \&
  {Gavriil}}]{2009ApJ...702..614D}
{Dib}, R., {Kaspi}, V.~M., \& {Gavriil}, F.~P. 2009, \apj, 702, 614

\bibitem[{{Duncan} \& {Thompson}(1992)}]{duncan92}
{Duncan}, R.~C. \& {Thompson}, C. 1992, \apjl, 392, L9

\bibitem[{{Durant} {et~al.}(2011){Durant}, {Kargaltsev}, \& {Pavlov}}]{durant2011}
{Durant}, M., {Kargaltsev}, O., \& {Pavlov}, G.~G. 2011, \apj, in press; arXiv:1108.3340

\bibitem[{{Eracleous} {et~al.}(1991){Eracleous}, {Patterson}, \&
  {Halpern}}]{1991ApJ...370..330E}
{Eracleous}, M., {Patterson}, J., \& {Halpern}, J. 1991, \apj, 370, 330

\bibitem[{{Esposito} {et~al.}(2010){Esposito}, {Israel}, {Turolla}, {Tiengo},
  {G{\"o}tz}, {de Luca}, {Mignani}, {Zane}, {Rea}, {Testa}, {Caraveo}, {Chaty},
  {Mattana}, {Mereghetti}, {Pellizzoni}, \& {Romano}}]{2010MNRAS.405.1787E}
{Esposito}, P., {Israel}, G.~L., {Turolla}, R., {et~al.} 2010, \mnras, 405,
  1787

\bibitem[{{Fahlman} \& {Gregory}(1981)}]{fahlman81}
{Fahlman}, G.~G. \& {Gregory}, P.~C. 1981, Nature, 293, 202

\bibitem[{{Ferrari} \& {Ruffini}(1969)}]{ferrari69}
{Ferrari}, A. \& {Ruffini}, R. 1969, \apjl, 158, L71+

\bibitem[{{Ferrario} {et~al.}(1997){Ferrario}, {Vennes}, {Wickramasinghe},
  {Bailey}, \& {Christian}}]{ferrario97}
{Ferrario}, L., {Vennes}, S., {Wickramasinghe}, D.~T., {Bailey}, J.~A., \&
  {Christian}, D.~J. 1997, \mnras, 292, 205

\bibitem[{{Ferrario} \& {Wickramasinghe}(2006)}]{ferrario06}
{Ferrario}, L. \& {Wickramasinghe}, D. 2006, \mnras, 367, 1323

\bibitem[{{Ferrario} \& {Wickramasinghe}(2005)}]{ferrario05}
{Ferrario}, L. \& {Wickramasinghe}, D.~T. 2005, \mnras, 356, 615

\bibitem[{{Ferrario} {et~al.}(2005){Ferrario}, {Wickramasinghe}, {Liebert}, \& {Williams}}]{ferrario05b}
{Ferrario}, L., {Wickramasinghe}, D., {Liebert}, J., \& {Williams}, K.~A. 2005, \mnras, 361, 1131

\bibitem[{{Gaensler} {et~al.}(2001){Gaensler}, {Slane}, {Gotthelf}, \& {Vasisht}}]{gaensler01}
{Gaensler}, B.~M, {Slane}, P.~O., {Gotthelf}, E.~V., \& {Vasisht}, G. 2001, \apj, 559, 963

\bibitem[{{Garcia-Berro} {et~al.}(1988){Garcia-Berro}, {Hernanz}, {Isern}, \&
  {Mochkovitch}}]{1988Natur.333..642G}
{Garcia-Berro}, E., {Hernanz}, M., {Isern}, J., \& {Mochkovitch}, R. 1988,
  Nature, 333, 642

\bibitem[{{Garcia-Berro} {et~al.}(2011){Garcia-Berro}, {Isern},
  {Loren-Aguilar}, {Rueda}, \& {Ruffini}}]{garcia2011}
{Garcia-Berro}, E., {Isern}, J., {Loren-Aguilar}, P., {Rueda}, J.~A., \&
  {Ruffini}, R. 2011, in preparation

\bibitem[{{Garc{\'{\i}}a-Berro} {et~al.}(2010){Garc{\'{\i}}a-Berro}, {Torres},
  {Althaus}, {Renedo}, {Lor{\'e}n-Aguilar}, {C{\'o}rsico}, {Rohrmann},
  {Salaris}, \& {Isern}}]{2010Natur.465..194G}
{Garc{\'{\i}}a-Berro}, E., {Torres}, S., {Althaus}, L.~G., {et~al.} 2010,
  Nature, 465, 194

\bibitem[{{Gelfand} \& {Gaensler}(2007)}]{2007ApJ...667.1111G}
{Gelfand}, J.~D. \& {Gaensler}, B.~M. 2007, \apj, 667, 1111

\bibitem[{{Giacconi}(2002)}]{giacconi02}
{Giacconi}, R. 2002, {The dawn of x-ray astronomy. Nobel lecture.}, ed.
  {Giacconi, R.}

\bibitem[{{Giacconi} \& {Ruffini}(1978 Reprinted 2010)}]{giacconi78gral}
{Giacconi}, R. \& {Ruffini}, R., eds. 1978 Reprinted 2010, {Physics and
  astrophysics of neutron stars and black holes}

\bibitem[{{Goldstein} {et~al.}(2011){Goldstein}, {Preece}, {Briggs}, {van der
  Horst}, {McBreen}, {Kouveliotou}, {Connaughton}, {Paciesas}, {Meegan},
  {Bhat}, {Bissaldi}, {Burgess}, {Chaplin}, {Diehl}, {Fishman}, {Fitzpatrick},
  {Foley}, {Gibby}, {Giles}, {Greiner}, {Gruber}, {Guiriec}, {von Kienlin},
  {Kippen}, {Rau}, {Tierney}, \& {Wilson-Hodge}}]{goldstein11}
{Goldstein}, A., {Preece}, R.~D., {Briggs}, M.~S., {et~al.} 2011, arXiv:1101.2458
  e-prints

\bibitem[{{Gonzalez} {et~al.}(2007){Gonzalez}, {Kaspi}, {Camilo}, {Gaensler},
  \& {Pivovaroff}}]{2007Ap&SS.308...89G}
{Gonzalez}, M.~E., {Kaspi}, V.~M., {Camilo}, F., {Gaensler}, B.~M., \&
  {Pivovaroff}, M.~J. 2007, Astroph. Space Sc., 308, 89

\bibitem[{{Gregory} \& {Fahlman}(1980)}]{gregory80}
{Gregory}, P.~C. \& {Fahlman}, G.~G. 1980, Nature, 287, 805

\bibitem[{{Hansen}(2002)}]{hansen02}
{Hansen}, B.~M.~S. 2002, in Astronomical Society of the Pacific Conference
  Series, Vol. 263, Stellar Collisions, Mergers and their Consequences, ed.
  {M.~M.~Shara}, 221

\bibitem[{{Hartle}(1967)}]{1967ApJ...150.1005H}
{Hartle}, J.~B. 1967, \apj, 150, 1005

\bibitem[{{Hartle} \& {Thorne}(1968)}]{1968ApJ...153..807H}
{Hartle}, J.~B. \& {Thorne}, K.~S. 1968, \apj, 153, 807

\bibitem[{{Hewish}(1974)}]{hewish74}
{Hewish}, A. 1974, {Pulsars and High Density Physics. Nobel lecture.}, ed.
  {Hewish, A.}

\bibitem[{{Hughes} {et~al.}(1981){Hughes}, {Harten}, \& {van den
  Bergh}}]{hughes81}
{Hughes}, V.~A., {Harten}, R.~H., \& {van den Bergh}, S. 1981, \apjl, 246, L127

\bibitem[{{Hulleman} {et~al.}(2000){Hulleman}, {van Kerkwijk}, \& {Kulkarni}}]{hulleman2000}
{Hulleman}, F., {van Kerkwijk}, M.~H., \& {Kulkarni}, S.~R. 2000, Nature, 408, 689

\bibitem[{{Hurley} {et~al.}(2005){Hurley}, {Boggs}, {Smith}, {Duncan}, {Lin},
  {Zoglauer}, {Krucker}, {Hurford}, {Hudson}, {Wigger}, {Hajdas}, {Thompson},
  {Mitrofanov}, {Sanin}, {Boynton}, {Fellows}, {von Kienlin}, {Lichti}, {Rau},
  \& {Cline}}]{2005Natur.434.1098H}
{Hurley}, K., {Boggs}, S.~E., {Smith}, D.~M., {et~al.} 2005, \nat, 434, 1098

\bibitem[{{Iben} \& {Tutukov}(1984)}]{iben84}
{Iben}, Jr., I. \& {Tutukov}, A.~V. 1984, \apjs, 54, 335

\bibitem[{{Ikhsanov} \& {Beskrovnaya}(2008)}]{ikhsanov08}
{Ikhsanov}, N.~R. \& {Beskrovnaya}, N.~G. 2008, arXiv:0809.1169

\bibitem[{{Ikhsanov} \& {Biermann}(2006)}]{ikhsanov06}
{Ikhsanov}, N.~R. \& {Biermann}, P.~L. 2006, \aap, 445, 305

\bibitem[{{Israel} {et~al.}(2007){Israel}, {Campana}, {Dall'Osso}, {Muno},
  {Cummings}, {Perna}, \& {Stella}}]{israel07}
{Israel}, G.~L., {Campana}, S., {Dall'Osso}, S., {et~al.} 2007, \apj, 664, 448

\bibitem[{{Israel} {et~al.}(1997){Israel}, {Stella}, {Angelini}, {White},
  {Kallman}, {Giommi}, \& {Treves}}]{1997ApJ...474L..53I}
{Israel}, G.~L., {Stella}, L., {Angelini}, L., {et~al.} 1997, \apjl, 474, L53+

\bibitem[{{Itoh} {et~al.}(2006){Itoh}, {Okada}, {Ishida}, \&
  {Kunieda}}]{2006ApJ...639..397I}
{Itoh}, K., {Okada}, S., {Ishida}, M., \& {Kunieda}, H. 2006, \apj, 639, 397

\bibitem[{{Judge} {et~al.}(2003){Judge}, {Solomon}, \& {Ayres}}]{judge03}
{Judge}, P.~G., {Solomon}, S.~C., \& {Ayres}, T.~R. 2003, \apj, 593, 534

\bibitem[{{Kasen} \& {Bildsten}(2010)}]{kasen10}
{Kasen}, D. \& {Bildsten}, L. 2010, \apj, 717, 245

\bibitem[{{Kashiyama} {et~al.}(2011){Kashiyama}, {Ioka}, \&
  {Kawanaka}}]{2011PhRvD..83b3002K}
{Kashiyama}, K., {Ioka}, K., \& {Kawanaka}, N. 2011, \prd, 83, 023002

\bibitem[{{Kaspi} {et~al.}(2003){Kaspi}, {Gavriil}, {Woods}, {Jensen},
  {Roberts}, \& {Chakrabarty}}]{kaspi03}
{Kaspi}, V.~M., {Gavriil}, F.~P., {Woods}, P.~M., {et~al.} 2003, \apjl, 588, L93

\bibitem[{{Katz}(1996){Katz}}]{1996ApJ...463..305K}
{Katz}, J.~I. 1996, \apj, 463, 305

\bibitem[{{Kepler} {et~al.}(2010){Kepler}, {Kleinman}, {Pelisoli}, {Pe{\c
  c}anha}, {Diaz}, {Koester}, {Castanheira}, \& {Nitta}}]{2010AIPC.1273...19K}
{Kepler}, S.~O., {Kleinman}, S.~J., {Pelisoli}, I., {et~al.} 2010, in American
  Institute of Physics Conference Series, Vol. 1273, American Institute of
  Physics Conference Series, ed. {K.~Werner \& T.~Rauch}, 19--24

\bibitem[{{Konacki} {et~al.}(1999){Konacki}, {Lewandowski}, {Wolszczan},
  {Doroshenko}, \& {Kramer}}]{konacki99}
{Konacki}, M., {Lewandowski}, W., {Wolszczan}, A., {Doroshenko}, O., \&
  {Kramer}, M. 1999, \apjl, 519, L81

\bibitem[{{Konacki} \& {Wolszczan}(2003)}]{konacki03}
{Konacki}, M. \& {Wolszczan}, A. 2003, \apjl, 591, L147

\bibitem[{{Kuiper} \& {Hermsen}(2009)}]{2009A&A...501.1031K}
{Kuiper}, L. \& {Hermsen}, W. 2009, \aap, 501, 1031

\bibitem[{{K{\"u}lebi} {et~al.}(2010{\natexlab{a}}){K{\"u}lebi}, {Jordan},
  {Euchner}, {Gaensicke}, \& {Hirsch}}]{2010yCat..35061341K}
{K{\"u}lebi}, B., {Jordan}, S., {Euchner}, F., {Gaensicke}, B.~T., \& {Hirsch},
  H. 2010{\natexlab{a}}, VizieR Online Data Catalog, 350, 61341

\bibitem[{{K{\"u}lebi} {et~al.}(2009){K{\"u}lebi}, {Jordan}, {Euchner},
  {G{\"a}nsicke}, \& {Hirsch}}]{2009A&A...506.1341K}
{K{\"u}lebi}, B., {Jordan}, S., {Euchner}, F., {G{\"a}nsicke}, B.~T., \&
  {Hirsch}, H. 2009, \aap, 506, 1341

\bibitem[{{K{\"u}lebi} {et~al.}(2010{\natexlab{b}}){K{\"u}lebi}, {Jordan},
  {Nelan}, {Bastian}, \& {Altmann}}]{2010A&A...524A..36K}
{K{\"u}lebi}, B., {Jordan}, S., {Nelan}, E., {Bastian}, U., \& {Altmann}, M.
  2010{\natexlab{b}}, \aap, 524, A36+

\bibitem[{{Kumar} \& {Safi-Harb}(2008)}]{2008ApJ...678L..43K}
{Kumar}, H.~S. \& {Safi-Harb}, S. 2008, \apjl, 678, L43

\bibitem[{{Levan} {et~al.}(2006){Levan}, {Wynn}, {Chapman}, {Davies}, {King},
  {Priddey}, \& {Tanvir}}]{levan06}
{Levan}, A.~J., {Wynn}, G.~A., {Chapman}, R., {et~al.} 2006, \mnras, 368, L1

\bibitem[{{Levin} {et~al.}(2010){Levin}, {Bailes}, {Bates}, {Bhat}, {Burgay},
  {Burke-Spolaor}, {D'Amico}, {Johnston}, {Keith}, {Kramer}, {Milia},
  {Possenti}, {Rea}, {Stappers}, \& {van Straten}}]{2010ApJ...721L..33L}
{Levin}, L., {Bailes}, M., {Bates}, S., {et~al.} 2010, \apjl, 721, L33

\bibitem[{{Liebert} {et~al.}(1983){Liebert}, {Schmidt}, {Green}, {Stockman}, \&
  {McGraw}}]{1983ApJ...264..262L}
{Liebert}, J., {Schmidt}, G.~D., {Green}, R.~F., {Stockman}, H.~S., \&
  {McGraw}, J.~T. 1983, \apj, 264, 262

\bibitem[{{Livingstone} {et~al.}(2010){Livingstone}, {Ng}, {Kaspi}, {Gavriil},
  \& {Gotthelf}}]{livingstone10}
{Livingstone}, M.~A., {Ng}, C., {Kaspi}, V.~M., {Gavriil}, F.~P., \&
  {Gotthelf}, E.~V. 2010, \apj, 264, 262
  
 \bibitem[{{Livingstone} {et~al.}(2010){Livingstone}, {Kaspi}, \& {Gavriil}}]{livingstone10b}
{Livingstone}, M.~A., {Kaspi}, V.~M., \& {Gavriil}, F.~P. 2010, \apj, 710, 1710  

\bibitem[{{Lyutikov} \& {Gavriil}(2006)}]{lyutikov06}
{Lyutikov}, M. \& {Gavriil}, F.~P. 2006, \mnras, 368, 690

\bibitem[{{Manchester} \& {Taylor}(1977)}]{manchesterbook}
{Manchester}, R.~N. \& {Taylor}, J.~H. 1977, {Pulsars}, {San Francisco, W. H. Freeman}

\bibitem[{{Mazets} {et~al.}(1979){Mazets}, {Golentskii}, {Ilinskii}, {Aptekar},
  \& {Guryan}}]{mazets79}
{Mazets}, E.~P., {Golentskii}, S.~V., {Ilinskii}, V.~N., {Aptekar}, R.~L., \&
  {Guryan}, I.~A. 1979, Nature, 282, 587

\bibitem[{{Meintjes} {et~al.}(1993){Meintjes}, {de Jager}, \& {et
  al.}}]{1993ICRC....1..338M}
{Meintjes}, P.~J., {de Jager}, C.~O., \& {et al.} 1993, in International Cosmic
  Ray Conference, Vol.~1, International Cosmic Ray Conference, 338--+

\bibitem[{{Meintjes} {et~al.}(1992){Meintjes}, {Raubenheimer}, {de Jager},
  {Brink}, {Nel}, {North}, {van Urk}, \& {Visser}}]{1992ApJ...401..325M}
{Meintjes}, P.~J., {Raubenheimer}, B.~C., {de Jager}, O.~C., {et~al.} 1992,
  \apj, 401, 325

\bibitem[{{Mereghetti}(2008)}]{mereghetti08}
{Mereghetti}, S. 2008, \aapr, 15, 225

\bibitem[{{Mereghetti} {et~al.}(2005){Mereghetti}, {Tiengo}, {Esposito},
  {G{\"o}tz}, {Stella}, {Israel}, {Rea}, {Feroci}, {Turolla}, \&
  {Zane}}]{2005ApJ...628..938M}
{Mereghetti}, S., {Tiengo}, A., {Esposito}, P., {et~al.} 2005, \apj, 628, 938

\bibitem[{{Mereghetti} {et~al.}(2009){Mereghetti}, {Tiengo}, {Esposito}, {La
  Palombara}, {Israel}, \& {Stella}}]{2009Sci...325.1222M}
{Mereghetti}, S., {Tiengo}, A., {Esposito}, P., {et~al.} 2009, Science, 325,
  1222
  
\bibitem[{{Mereghetti} {et~al.}(2011){Mereghetti}, {La
  Palombara}, {Tiengo}, {Pizzolato}, {Esposito}, {Woudt}, {Israel}, \& {Stella}}]{mereghettiwd2011}
{Mereghetti}, S., {La Palombara}, N., {Tiengo}, A., {Pizzolato}, F., {Esposito}, P., {Woudt}, P.~A., {Israel}, G.~L.,  {et~al.}, \& {Stella}, L. 2011, \apj, 737, 51

\bibitem[{{Michel} \& {Dessler}(1981)}]{michel81}
{Michel}, F.~C \& {Dessler}, A.~J. 1981, \apj, 251, 664 

\bibitem[{{Michel}(1981)}]{michel83}
{Michel}, F.~C 1983, \apj, 266, 188 

\bibitem[{{Morini} {et~al.}(1988){Morini}, {Robba}, {Smith}, \& {van der
  Klis}}]{1988ApJ...333..777M}
{Morini}, M., {Robba}, N.~R., {Smith}, A., \& {van der Klis}, M. 1988, \apj,
  333, 777

\bibitem[{{Nale{\.z}yty} \& {Madej}(2004)}]{nalezyty04}
{Nale{\.z}yty}, M. \& {Madej}, J. 2004, \aap, 420, 507

\bibitem[{{Ng} {et~al.}(2010){Ng}, {Kaspi}, {Dib}, {Olausen}, {Scholz},
  {Guver}, {Ozel}, {Gavriil}, \& {Woods}}]{ng10}
{Ng}, C., {Kaspi}, V.~M., {Dib}, R., {et~al.} 2011, \apj, 729, 131

\bibitem[{{Ostriker} \& {Bodenheimer}(1968)}]{1968ApJ...151.1089O}
{Ostriker}, J.~P. \& {Bodenheimer}, P. 1968, \apj, 151, 1089

\bibitem[{{Paczynski}(1985)}]{paczynski85}
{Paczynski}, B. 1985, in Astrophysics and Space Science Library, Vol. 113,
  Cataclysmic Variables and Low-Mass X-ray Binaries, ed. {D.~Q.~Lamb \&
  J.~Patterson}, 1--12

\bibitem[{{Paczynski}(1990)}]{paczynski90}
{Paczynski}, B. 1990, \apjl, 365, L9

\bibitem[{{Paczynski}(1992)}]{paczynski92}
{Paczynski}, B. 1992, Acta Astronomica, 42, 145

\bibitem[{{Pakmor} {et~al.}(2010){Pakmor}, {Kromer}, {R{\"o}pke}, {Sim},
  {Ruiter}, \& {Hillebrandt}}]{pakmor10}
{Pakmor}, R., {Kromer}, M., {R{\"o}pke}, F.~K., {et~al.} 2010, Nature, 463, 61

\bibitem[{{Parker}(1957)}]{parker57}
{Parker}, E.~N. 1957, Physical Review, 107, 830

\bibitem[{{Patnaude} {et~al.}(2009){Patnaude}, {Loeb}, \& {Jones}}]{patnaude09}
{Patnaude}, D.~J., {Loeb}, A., \& {Jones}, C. 2009, arXiv:0912.1571

\bibitem[{{Peres} {et~al.}(2000){Peres}, {Orlando}, {Reale}, {Rosner}, \&
  {Hudson}}]{peres2000}
{Peres}, G., {Orlando}, S., {Reale}, F., {Rosner}, R., \& {Hudson}, H. 2000,
  \apj, 528, 537

\bibitem[{{Rea} {et~al.}(2010){Rea}, {Esposito}, {Turolla}, {Israel}, {Zane},
  {Stella}, {Mereghetti}, {Tiengo}, {G{\"o}tz}, {G{\"o}{\u g}{\"u}{\c s}}, \&
  {Kouveliotou}}]{rea10}
{Rea}, N., {Esposito}, P., {Turolla}, R., {et~al.} 2010, Science, 330, 944

\bibitem[{{Rea} {et~al.}(2011){Rea}, {Jonker}, {Nelemans}, {Pons}, {Kasliwal},
  {Kulkarni}, \& {Wijnands}}]{rea11}
{Rea}, N., {Jonker}, P.~G., {Nelemans}, G., {et~al.} 2011, \apj, 729, L21

\bibitem[{{Rea} {et~al.}(2008){Rea}, {Zane}, {Turolla}, {Lyutikov}, \&
  {G{\"o}tz}}]{rea08}
{Rea}, N., {Zane}, S., {Turolla}, R., {Lyutikov}, M., \& {G{\"o}tz}, D. 2008, 
\apj, 686, 1245

\bibitem[{{Romani} \& {Watters}(2010)}]{romani2010}
{Romani}, Roger W. \& {Watters}, Kyle P. 2010, \apj, 714, 810--824

\bibitem[{{Rotondo} {et~al.}(2011){Rotondo}, {Rueda}, {Ruffini}, \&
  {Xue}}]{2011PhRvC..83d5805R}
{Rotondo}, M., {Rueda}, J.~A., {Ruffini}, R., \& {Xue}, S.-S. 2011, \prc, 83,
  045805

\bibitem[{{Rotondo} {et~al.}(2011){Rotondo}, {Rueda}, {Ruffini}, \&
  {Xue}}]{rueda10wd}
{Rotondo}, M., {Rueda}, J.~A., {Ruffini}, R., \& {Xue}, S. 2011, \prd, 84, 084007

\bibitem[{{Rueda} {et~al.}(2011){Rueda}, {Ruffini}, \& {Xue}}]{ruedaNPA}
{Rueda}, J.~A., {Ruffini}, R., \& {Xue}, S. 2011, Nucl.\ Phys.\ A, in press; arXiv:1104.4062

\bibitem[{{Ruffini}(1973)}]{ruffini73}
{Ruffini}, R. 1973, in Black Holes (Les Astres Occlus), ed. {A.~Giannaras},
  451--546

\bibitem[{{Ruffini}(2009)}]{ruffinimg12}
{Ruffini}, R. 2009, in Proceedings of the 12th Marcel Grossmann Meeting on
  General Relativity, ed. {T.~Damour and R.~Ruffini}, World Scientific,
  Singapore

\bibitem[{{Sasaki} {et~al.}(2004){Sasaki}, {Plucinsky}, {Gaetz}, {Smith},
  {Edgar}, \& {Slane}}]{sasaki04}
{Sasaki}, M., {Plucinsky}, P.~P., {Gaetz}, T.~J., {et~al.} 2004, \apj, 617, 322

\bibitem[{{Schmidt} {et~al.}(1992){Schmidt}, {Bergeron}, {Liebert}, \&
  {Saffer}}]{1992ApJ...394..603S}
{Schmidt}, G.~D., {Bergeron}, P., {Liebert}, J., \& {Saffer}, R.~A. 1992, \apj,
  394, 603

\bibitem[{{Schmidt} {et~al.}(1986){Schmidt}, {West}, {Liebert}, {Green}, \&
  {Stockman}}]{1986ApJ...309..218S}
{Schmidt}, G.~D., {West}, S.~C., {Liebert}, J., {Green}, R.~F., \& {Stockman},
  H.~S. 1986, \apj, 309, 218

\bibitem[{{Shapiro} \& {Teukolsky}(1983)}]{shapirobook}
{Shapiro}, S.~L. \& {Teukolsky}, S.~A. 1983, {Black holes, white dwarfs, and
neutron stars: The physics of compact objects}, {New York, Wiley-Interscience}

\bibitem[{{Stairs}(2004)}]{2004Sci...304..547S}
{Stairs}, I.~H. 2004, Science, 304, 547

\bibitem[{{Stefanescu} {et~al.}(2008){Stefanescu}, {Kanbach}, {S{\l}owikowska},
  {Greiner}, {McBreen}, \& {Sala}}]{stefanescu08}
{Stefanescu}, A., {Kanbach}, G., {S{\l}owikowska}, A., {et~al.} 2008, Nature,
  455, 503

\bibitem[{{Sweet}(1958)}]{sweet58}
{Sweet}, P.~A. 1958, in IAU Symposium, Vol.~6, Electromagnetic Phenomena in
  Cosmical Physics, ed. {B.~Lehnert}, 123--+

\bibitem[{{Tauris} \& {Bailes}(1996)}]{1996A&A...315..432T}
{Tauris}, T.~M. \& {Bailes}, M. 1996, \aap, 315, 432

\bibitem[{{Tauris} \& {Takens}(1998)}]{1998A&A...330.1047T}
{Tauris}, T.~M. \& {Takens}, R.~J. 1998, \aap, 330, 1047

\bibitem[{{Terada}(2008)}]{2008AstHe.101..526T}
{Terada}, Y. 2008, Astronomical Herald, 101, 526

\bibitem[{{Terada} {et~al.}(2008{\natexlab{a}}){Terada}, {Hayashi}, {Ishida},
  {Makishima}, {Mukai}, {Dotani}, {Okada}, {Nakamura}, {Naik}, {Bamba}, \&
  {Morigami}}]{2008HEAD...10.1003T}
{Terada}, Y., {Hayashi}, T., {Ishida}, M., {et~al.} 2008{\natexlab{a}}, in
  AAS/High Energy Astrophysics Division, Vol.~10, AAS/High Energy Astrophysics
  Division \#10, 10.03--+

\bibitem[{{Terada} {et~al.}(2008{\natexlab{b}}){Terada}, {Hayashi}, {Ishida},
  {Mukai}, {Dotani}, {Bamba}, {Okada}, {Nakamura}, {Makishima}, {Morigami}, \&
  {Harayama}}]{2008AIPC.1085..689T}
{Terada}, Y., {Hayashi}, T., {Ishida}, M., {et~al.} 2008{\natexlab{b}}, in
  American Institute of Physics Conference Series, Vol. 1085, American
  Institute of Physics Conference Series, ed. {F.~A.~Aharonian, W.~Hofmann, \&
  F.~Rieger}, 689--692

\bibitem[{{Terada} {et~al.}(2008{\natexlab{c}}){Terada}, {Hayashi}, {Ishida},
  {Mukai}, {Dotani}, {Okada}, {Nakamura}, {Naik}, {Bamba}, \&
  {Makishima}}]{terada08}
{Terada}, Y., {Hayashi}, T., {Ishida}, M., {et~al.} 2008{\natexlab{c}}, Publ.
  Astron. Soc. Japan, 60, 387

\bibitem[{{Terada} {et~al.}(2008{\natexlab{d}}){Terada}, {Ishida}, {Mukai},
  {Dotani}, {Makishima}, {Naik}, {Hayashi}, {Okada}, {Nakamura}, \&
  {Enoto}}]{2008AdSpR..41..512T}
{Terada}, Y., {Ishida}, M., {Mukai}, K., {et~al.} 2008{\natexlab{d}}, Advances
  in Space Research, 41, 512

\bibitem[{{Thompson} \& {Duncan}(1995)}]{thompson95}
{Thompson}, C. \& {Duncan}, R.~C. 1995, \mnras, 275, 255

\bibitem[{{Thompson} {et~al.}(2002){Thompson}, {Lyutikov}, \&
  {Kulkarni}}]{thompson02}
{Thompson}, C., {Lyutikov}, M., \& {Kulkarni}, S.~R. 2002, \apj, 574, 332

\bibitem[{{Tiengo} {et~al.}(2005){Tiengo}, {Esposito}, {Mereghetti}, {Rea},
  {Stella}, {Israel}, {Turolla}, \& {Zane}}]{2005A&A...440L..63T}
{Tiengo}, A., {Esposito}, P., {Mereghetti}, S., {et~al.} 2005, \aap, 440, L63

\bibitem[{{Tong} {et~al.}(2010){Tong}, {Song}, \& {Xu}}]{tong10}
{Tong}, H., {Song}, L.~M., \& {Xu}, R.~X. 2010, \apjl, 725, L196

\bibitem[{{Tong} {et~al.}(2011){Tong}, {Song}, \& {Xu}}]{tong11}
{Tong}, H., {Song}, L.~M., \& {Xu}, R.~X. 2011, arXiv:1110.1975

\bibitem[{{Usov}(1994)}]{usov94}
{Usov}, V.~V. 1994, \apj, 427, 984

\bibitem[{{van Paradijs} {et~al.}(1995){van Paradijs}, {Taam}, \& {van den
  Heuvel}}]{1995A&A...299L..41V}
{van Paradijs}, J., {Taam}, R.~E., \& {van den Heuvel}, E.~P.~J. 1995, \aap,
  299, L41+

\bibitem[{{Vink}(2008)}]{vink08}
{Vink}, J. 2008, Advances in Space Research, 41, 503

\bibitem[{{Vink} \& {Kuiper}(2006)}]{vink06}
{Vink}, J. \& {Kuiper}, L. 2006, \mnras, 370, L14

\bibitem[{{Woods} {et~al.}(2004){Woods}, {Kaspi}, {Thompson}, {Gavriil},
  {Marshall}, {Chakrabarty}, {Flanagan}, {Heyl}, \& {Hernquist}}]{woods04}
{Woods}, P.~M., {Kaspi}, V.~M., {Thompson}, C., {et~al.} 2004, \apj, 605, 378

\bibitem[{{Woosley}(2010)}]{woosley10}
{Woosley}, S.~E. 2010, \apjl, 719, L204

\bibitem[{{Xu} {et~al.}(2006){Xu}, {Tao}, \& {Yang}}]{xu06}
{Xu}, R.~X., {Tao}, D.~J., \& {Yang}, Y. 2006, \mnras, 373, L85

\bibitem[{{Zane} {et~al.}(2009){Zane}, {Rea}, {Turolla}, \& {Nobili}}]{zane09}
{Zane}, S., {Rea}, N., {Turolla}, R., \& {Nobili}, L. 2009, \mnras, 398, 1403

\bibitem[{{Bing Zhang} \& {Janusz Gil}(2005)}]{zhang05}
{Zhang}, B., \& {Gil}, J. 2005, \apjl, 631, 143

\bibitem[{{Zhu} {et~al.}(2011){Zhu}, {Kaspi}, {McLaughlin}, {Pavlov}, {Ng},
  {Manchester}, {Gaensler}, \& {Woods}}]{2011ApJ...734...44Z}
{Zhu}, W.~W., {Kaspi}, V.~M., {McLaughlin}, M.~A., {et~al.} 2011, \apj, 734, 44

\end{thebibliography}
\end{document}